\documentclass[aps,pra,twocolumn,superscriptaddress,showpacs,longbibliography]{revtex4-1}
\usepackage{amsmath,amsfonts,amssymb,amsthm}
\usepackage{hyperref}
\usepackage{bbm}
\usepackage{tabularx}
\usepackage{physics}
\usepackage{titlesec}

\makeatletter
\renewcommand\p@subsection{\thesection.}
\makeatother

\titleclass{\subsubsubsection}{straight}[\subsection]
\newcounter{subsubsubsection}[subsubsection]
\renewcommand\thesubsubsubsection{\roman{subsubsubsection}.}
\titleformat{\subsubsubsection}{\normalsize\it}{\thesubsubsubsection}{1em}{}
\titlespacing*{\subsubsubsection}{0pt}{3ex}{1ex}

\makeatletter
\renewcommand\paragraph{%
	\@startsection {paragraph}{5}{\z@ }{3.25ex \@plus 1ex \@minus .2ex}{-1em}{\normalsize \it}}
\makeatother

\newcommand{\Bra}[1]{\left<\!\left< #1 \right.\right|}
\newcommand{\Ket}[1]{\left|\left. #1 \right>\!\right>}
\newcommand{\Braket}[2]{\left<\!\left< #1 \vphantom{#2} \right.\right|\hspace{-0.6ex}\left.\left. #2 \vphantom{#1}\right>\!\right>}
\newcommand{\PT}{${\cal PT}$}
\begin{document}
\title{Non-Hermitian Hamiltonians and no-go theorems in quantum 
information}

\author{Chia-Yi Ju}
\affiliation{Department of Physics, National Chung Hsing University, Taichung 402, Taiwan}
\author{Adam Miranowicz}
\affiliation{Faculty of Physics, Adam Mickiewicz University, 61-614 Pozna\'{n}, Poland}
\affiliation{Theoretical Quantum Physics Laboratory, RIKEN Cluster for Pioneering Research, Wako-shi, Saitama 351-0198, Japan}
\author{Guang-Yin Chen}
\email{gychen@phys.nchu.edu.tw}
\affiliation{Department of Physics, National Chung Hsing University, Taichung 402, Taiwan}
\author{Franco Nori}
\affiliation{Theoretical Quantum Physics Laboratory, RIKEN Cluster for Pioneering Research, Wako-shi, Saitama 351-0198, Japan}
\affiliation{Department of Physics, University of Michigan, Ann Arbor, Michigan 48109-1040, USA}

\begin{abstract}
	Recently, apparent nonphysical implications of non-Hermitian quantum mechanics (NHQM) have been discussed in the literature. In particular, the apparent violation of the no-signaling theorem, discrimination of nonorthogonal states, and the increase of quantum entanglement by local operations were reported, and therefore NHQM was not considered as a fundamental theory. Here we show that these and other no-go principles (including the no-cloning and no-deleting theorems) of conventional quantum mechanics still hold in finite-dimensional non-Hermitian quantum systems, including parity-time symmetric and pseudo-Hermitian cases, if its formalism is properly applied. We have developed a modified formulation of NHQM based on the geometry of Hilbert spaces which is consistent with the conventional quantum mechanics for Hermitian systems. Using this formulation the validity of these principles can be shown in a simple and uniform approach. 

\end{abstract}

\maketitle%

\newpage
	\section{Introduction \label{Introduction}}
		
		It is well established that quantum states reside in the corresponding Hilbert spaces and the time evolution of these states is governed by Schr\"{o}dinger's equation together with the Hamiltonian of the system. These Hamiltonians were considered to be Hermitian [note that throughout this paper, the word ``Hermitian'' is limited to the standard quantum mechanics (QM) textbook Hermitian] so that the eigenvalues can all be real.
		
		However, in 1998, Bender and Boettcher~\cite{BenderRealEigenvalues} discovered that the eigenvalues of the Hamiltonian for parity-inversion plus time-reversal symmetric (\PT-symmetric) systems are all real. This has attracted considerable attention on non-Hermitian physics, both theoretical and experimental, in several nonquantum areas ranging from classical optics, optomechanics, mechanics, acoustics, electronics, plasmonics, metamaterials, and photonic crystals to innovative devices (for references see, e.g., \cite{Peng2014, Jing2014, Jing2015, Zhang2015, Liu2017, Quijandria2018, Bliokh2019, Bliokh2019a} and a very recent review~\cite{OzdemirParity}). In contrast to many experiments with classical \PT-symmetric systems, the first experimental observation (or, rather, experimental simulation)~\cite{Xiao2018} of critical phenomena during non-unitary quantum dynamics of a \PT-symmetric system has been reported only very recently. We note that even the theoretical studies of exceptional points of \PT-symmetric systems~\cite{OzdemirParity} have been almost exclusively limited (except Refs.~\cite{Minganti2019, Arkhipov2019, Hatano2019}) to the semiclassical and classical regimes, where quantum jumps are completely ignored. Here, we study non-Hermitian quantum mechanics (NHQM) in the fully quantum regime. Of course, various fundamental aspects of the quantum regime of NHQM systems have already been thoroughly theoretically investigated (for comprehensive reviews of the theory see, e.g., \cite{Scholtz1992, BenderMakingSense, Mostafazadeh2010, Bagarello2015, ZnojilNonHermitianInteraction, Bender2019}). However, to our knowledge, the most popular quantum-information no-go theorems (except the no-signaling theorem) have not been explicitly proved for NHQM. These theorems include quantum no-cloning and quantum no-deleting, as well as the limitations of some kinds of quantum entanglement manipulation (see Table \ref{NoGoTable}). Of course, in the classical regime of NHQM systems, perfect cloning and deleting are allowed, and quantum entanglement disappears, so there is no merit to discuss these no-go theorems in this regime.
		
		Rather than being just a theoretical product, many experimental realizations of \PT-symmetric systems have been demonstrated, e.g., \cite{RuterPTObservation, SchindlerPTExperimental, ZhangPTOptical}. Besides \PT-symmetric systems, some further studies~\cite{BittnerPTBreaking, LiuMetrology, HodaeiEnhancedSensitivity, OzdemirParity} approached more general non-Hermitian Hamiltonians by perturbing the system around a \PT-symmetric system. Additional interesting physical phenomena~\cite{HeissPT, WiersigEnhancingSensitivity, PengLoss, Gao2015, JingHigh, Leykam2017, LuExceptional, ChenExceptional, ZhangPhonon, Liu2019, Ge2019} were also found in other non-Hermitian systems.
		
		Applying conventional quantum mechanics (CQM) directly on NHQM might lead to apparent violations of some no-go theorems~\cite{LeeYesSignaling, ChenIncreasingEntanglement,  PatiViolateInvariance, ZnojilPTFoundamental}, including the no-signaling, no-cloning, no-broadcasting, and no-deleting principles. These principles are closely related no-go theorems of fundamental importance in quantum physics, including especially quantum information. The inclusion of damping and amplification for NHQM systems, which results in mixing pure states, is by no means trivial even for standard QM. We note that, e.g., the no-deleting theorem for mixed states has not been proved or even properly formulated in neither standard QM nor NHQM.
	
		The violation of any of these no-go theorems would have enormous implications in physics and would lead to fundamental paradoxes. Thus, verifying whether a given quantum theory does \emph{not} violate these theorems could be a simple test of its physical validity. Recently, it was ``demonstrated'' that local parity-time symmetry (\PT symmetry) in the initial formalism~\cite{BenderRealEigenvalues} \emph{apparently} allows the perfect discrimination of nonorthogonal states and the violation of the no-signaling principle~\cite{LeeYesSignaling}. That work~\cite{LeeYesSignaling} concluded that ``this shows that the \PT-symmetric theory is either a trivial extension or likely false as a fundamental theory'' (in the Abstract of~\cite{LeeYesSignaling}), and ``Finally, while in our view these results essentially destroy any hope of \PT-symmetric quantum theory as a fundamental theory of nature, it could still be useful as an effective model or as a purely mathematical problem solving device'' (in the Conclusions of~\cite{LeeYesSignaling}).
		
		We note that the perfect discrimination of nonorthogonal states would also imply the violation of the no-cloning principle. As shown by Znojil~\cite{ZnojilPTFoundamental} (see also Brody~\cite{BrodyConsistency}), this apparent violation of the no-signaling principle results from ``an unfortunate use of one of the simplest but still inadequate, manifestly unphysical Hilbert spaces.'' In contrast to such a wrong approach, a proper choice of the Hilbert space does \emph{not} lead to any violation of the no-signaling principle~\cite{ZnojilPTFoundamental, BrodyConsistency}.
		
		Brody~\cite{BrodyConsistency} has shown that indeed various claims made by various authors on the violation of causality in some non-Hermitian quantum systems are not valid when the theory is formulated in an appropriate manner. The fully quantum regime of NHQM systems has been thoroughly investigated (for comprehensible reviews of the theory see, e.g., \cite{Scholtz1992, BenderMakingSense, Mostafazadeh2010, Bagarello2015, ZnojilNonHermitianInteraction, Bender2019}). However, to our knowledge the most popular quantum-information no-go theorems (except the no-signaling theorem) have not been proved yet for NHQM.
		
		Here we provide the concrete proofs of some no-go theorems which are especially important for quantum information and quantum communication. To our knowledge these no-go theorems have not yet been proved in detail for general NHQM. We note that causality in standard non-relativistic or even relativistic QM and the no-signaling constraints are inequivalent principles although related.
		
		Analogously, the no-signaling, no-cloning, and no-deleting theorems are closely related, but \emph{not} equivalent. For example, Horodecki and Ramanathan in their recent work~\cite{Horodecki2019} on relativistic causality and no-signaling paradigm for multipartite correlations in general physical theories showed that ``while the usual no-signaling constraints are sufficient, in general they are not necessary to ensure that a theory does not violate causality. ... causality only imposes a subset of the usual no-signaling conditions.''
		
		The main cause of the inconsistencies is clearly the inner product between states, as pointed out in \cite{ZnojilPTFoundamental} and \cite{BrodyConsistency}. There are some modifications to the inner products proposed in the literature which either abandoned the relation between QM and Hilbert space~\cite{MoiseyevCProduct, MoiseyevFProduct, DasInnerProduct} or were limited to some special cases of non-Hermitian systems~\cite{MostafazadehPsuedoHermitian, BenderComplex, BenderCPT, LiInnerProduct, BrodyBiorthogonal}. To find a general inner product that also preserves the notion of Hilbert space, we treat the Schr\"{o}dinger equation as a covariant derivative and find a compatible metric. The compatible metric is far from unique, but they are all subject to the same equation of motion that was found in Refs.~\cite{MostafazadehMetric, BilaAdiabatic, ZnojilNonHermitianInteraction} in various contexts. We also show that the choice of metrics is equivalent to choosing a preference of bases. The restriction on the transition functions between different choices of bases are also derived in this paper. Note that the metrics in Hermitian systems can always be chosen to be the identity so that the inner products are the same as the conventional ones. It can also be shown that in pseudo-Hermitian systems, including the ``charge''-parity-time ($\cal{CPT}$) systems, the metric can be chosen to be the standard pseudo-Hermitian metric~\cite{MostafazadehPsuedoHermitian}, which is time independent. Furthermore, when the eigenstates of the Hamiltonian form a complete set of bases, with a special choice of the metric, biorthogonal quantum mechanics~\cite{BrodyBiorthogonal}  (BQM) can be recovered.
		
		In this paper, we focus on systems with finite-dimensional state space and demonstrate that a correct application of the NHQM formalism, using a proper Hilbert-space metric, together with the generalized operators derived in Sec.~\ref{GeneralizedOperators}, does \emph{not} lead to any violations of the principles studied in Refs.~\cite{LeeYesSignaling, ChenIncreasingEntanglement, PatiViolateInvariance}. Moreover, we show that the NHQM does \emph{not} violate other no-go theorems, including the no-cloning and no-deleting principles.
		
		One can raise the objection about the merit of proving the quantum-information no-go theorems for NHQM. Indeed, the consistency of NHQM seems widely accepted at present; thus, one can argue that such theorems cannot be violated by default. However, we note that although the consistency of conventional Hermitian QM has been accepted for a century, the listed six no-go theorems have been formulated and proved in CQM only in the 1980s and 1990s. Arguably, these theorems have triggered impressive interest and progress in quantum information, which shows the merit of proving the no-go theorems also for quantum information processing with NHQM systems.

	\section{Generalized Inner Product in NHQM \label{InnerProductScetion}}
		
		In QM, the probability of a state is the norm squared of the corresponding vector in a Hilbert space. The dynamics of the vector $\ket{ \psi(t) }$ has to satisfy the Schr\"{o}dinger equation,
		\begin{align}
			i \hbar \partial_t \ket{ \psi (t) } = H(t) \ket { \psi (t)}. \label{SchrodingersEquation}
		\end{align}
		For convenience, we refer to $H$ as a Hamiltonian, but it should be understood as a generalized Hamiltonian-type operator. If the Hamiltonian $H(t)$ is Hermitian, then the norm squared is conserved in time, because there is an obvious symmetry of the Hamiltonian [${H(t) = H^\dagger (t)}$], so that the inner product is not hard to find. On the other hand, for a Hamiltonian that does not have such a clear symmetry, the inner product that preserves the norm squared, is not so easy to find. However, Eq.~\eqref{SchrodingersEquation}, if written as
		\begin{align}
			\nabla_t \ket{ \psi (t) } = \left( \partial_t + \frac{i}{\hbar} H(t) \right) \ket{ \psi (t) } = 0,
		\end{align}
		suggests that $\nabla_t$ plays the role of a connection{~\cite{NakaharaFiberBundle}} in a vector bundle which connects the ``geometries'' of nearby points and $\ket{ \psi (t) }$ is therefore parallel transported along the time direction (moving the vector along the time curve while keeping it parallel to itself in different geometries along the curve). This analogy hints that there is a ``connection-compatible metric'' such that the inner products between the parallel transported vectors are time independent. Hence, the norm squared, defined as the inner product of the vector and itself, is also time independent.
		
		To distinguish the modified inner product from the conventional one, the modified inner product is denoted as ${\Braket{\psi_1 (t)}{\psi_2 (t)}}$ (see Table~\ref{TableI}). Note that there is no distinction between the new vector [$\Ket{\psi (t)}$], and the conventional ones [$\ket{\psi (t)}$] since both evolve according to the Schr\"{o}dinger equation. However, the dual vectors are not just the Hermitian conjugate of the conventional vectors, but they are also subject to a linear map, i.e.,
		\begin{align}
			\Bra{\psi (t)} = \bra{\psi(t)} G(t),
		\end{align}
		where $\bra{\psi(t)}$ is the standard Hermitian conjugate of $\ket{\psi(t)}$ and $G(t)$ plays a similar role of a fiber metric~\cite{NakaharaFiberBundle}. Therefore, hereafter $G(t)$ will be named the \emph{metric operator} or, simply, the \emph{metric}.
		
		\subsection{Metric}
		
			For reasons that will be discussed shortly, the metric $G(t)$ has to be Hermitian, positive-definite, and satisfying the equation of motion (see Table~\ref{TableI}):
			\begin{align}
				\partial_t G(t) = \frac{i}{\hbar} \left[ G(t) H (t) - H^\dagger (t) G(t)\right].  \label{MetricEquation}
			\end{align}
			
			At first, a time-dependent metric might seem peculiar, because naively this would require a proper reference point of time to begin with. However, since the norm squared is time invariant, any reference time works the same and the physics does not depend on it. In fact, metrics do not always have to be time dependent, because in many cases the corresponding metric can be time independent; for example, for the Hermitian cases~\cite{DiracQuantumMechanics}, $\mathcal{CPT}$-symmetric cases~\cite{BenderCPT,BenderComplex,BenderMakingSense}, and pseudo-Hermitian cases~\cite{MostafazadehPsuedoHermitian}. This can be easily seen by taking the time derivative in Eq.~\eqref{MetricEquation} to zero so that it reduces to the definition of pseudo-Hermitian, ${G H = H^\dagger G}$.
			
			The reasons why the metric needs to satisfy the above-mentioned constraints are as follows: For the probability to be time invariant, the time derivative on the inner product of an arbitrary vector with itself should vanish,
			\begin{align}
				\begin{split}
					& 0 = \partial_t \Braket{\psi (t)}{\psi (t)}\\
					& = \bra{\psi (t)} \left[ \partial_t G(t) + \frac{i}{\hbar} H^\dagger (t) G(t) - \frac{i}{\hbar} G(t) H (t) \right] \ket{\psi (t)}.
				\end{split}
			\end{align}
			Therefore, Eq.~\eqref{MetricEquation} is a necessary condition for the probability to be conserved.
	
			Furthermore, it should be obvious that the Hermitian conjugate of the metric, $G^\dagger (t)$, has to satisfy the same equation as for $G(t)$; therefore, the metric can and should be chosen to be Hermitian so that
			\begin{align}
				\begin{split}
					& \overline{\Braket{\psi_1 (t)}{\psi_2 (t)}}= \left[ \bra{\psi_1 (t)} G(t) \ket{\psi_2 (t)}\right]^\dagger\\
				 & \quad = \bra{\psi_2 (t)} G^\dagger (t) \ket{\psi_1 (t)} = \Braket{\psi_2 (t)}{\psi_1 (t)}.
				 \end{split}\label{ComplexConjugate}
			\end{align}
			Note that this is true only when the bra and ket vectors are at the same instant.
			
			Since the constraint, given in Eq.~\eqref{MetricEquation}, is a differential equation, there are some undetermined constant(s) in the solution. These constants can always be chosen so that $G(t)$ is positive definite,
			\begin{align}
				\Braket{\psi (t)}{\psi (t)} = \bra{\psi (t)} G(t) \ket{\psi (t)} > 0,
			\end{align}
			for every nonzero $\ket{\psi (t)}$. The positive definiteness and linearity of the Hermitian metric together with Eq.~\eqref{ComplexConjugate} is sufficient for the space equipped with this inner product to be a Hilbert space (assuming that the space is complete)~\cite{AppelPreHilbertSpace}.
			
			Note that despite that this formalism is formally correct, Eq.~\eqref{MetricEquation} is not always guaranteed to have a solution for every infinite-dimensional $H(t)$. There might need to be some additional constraints to rule out the unphysical solutions, e.g., divergent norms. Therefore, we focus only on the systems with finite-dimensional Hilbert spaces in the remainder of this paper.
			
			\subsection{Different Metrics as Different Bases}
			
			Although $G(t)$ is not uniquely determined for a given $H(t)$, different metrics $G(t)$ are related by a covariantly constant transition function. That is to say, given a Hamiltonian, if $G_1(t)$ and $G_2(t)$ are two possible metrics, there exists a function $T_{12}$ such that
			\begin{align}
				G_2(t) = T_{12}^\dagger (t) G_1(t) T_{12}(t),
			\end{align}
			where $T_{12}(t)$ satisfies
			\begin{align}
				\partial_t T_{12}(t) + \frac{i}{\hbar}  H (t) T_{12}(t) - \frac{i}{\hbar} T_{12}(t) H (t) = 0.
			\end{align}
			In other words, different choices of a metric $G(t)$ correspond to different choices of bases, and, therefore, are physically equivalent. That is to say, although the two metrics may look different, as long as the bases and the operators are also transformed by the same transition function $T_{12}(t)$, the physical contents will not be altered.
			
			\begin{table*}[t]
				\renewcommand*{\arraystretch}{1.6}
				\begin{tabular}{| >{\centering\arraybackslash}m{4cm} | >{\centering\arraybackslash}m{6cm} | >{\centering\arraybackslash}m{6cm} |}
					\hline
					& Conventional quantum mechanics & Non-Hermitian quantum mechanics\\
					\hline
					Equation(s) of motion & $\displaystyle i \hbar \partial_t \ket{\psi(t)} = H \ket{\psi(t)}$ & \begin{tabular}{c} $\displaystyle i \hbar \partial_t \Ket{\psi(t)} = H \Ket{\psi(t)}$,\\ $\displaystyle -i \hbar \partial_t G(t) = G(t) H (t) - H^\dagger (t) G(t)$\end{tabular}\\
					\hline
					Inner product & $\braket{\psi(t)}{\phi(t)}$& $\Braket{\psi(t)}{\phi(t)} = \bra{\psi(t)}G(t)\ket{\phi(t)}$\\
					\hline
					Complex conjugation & $\overline{\braket{\psi(t)}{\phi(t)}} = \braket{\phi(t)}{\psi(t)}$ & $\overline{\Braket{\psi(t)}{\phi(t)}} = \Braket{\phi(t)}{\psi(t)}$\\
					\hline
					Completeness relation & $\displaystyle \sum_n \ket{n(t)} \bra{n(t)} = \mathbbm{1}$ & $\displaystyle \sum_n \Ket{n(t)} \Bra{n(t)} = \mathbbm{1}$\\
					\hline
				\end{tabular}
				\caption{Some differences between conventional quantum mechanics, CQM (left), and the non-trivial metric one (right) for non-Hermitian quantum mechanics.\label{TableI}}
			\end{table*}		
			
			In fact, if ${\lbrace \Ket{n(t)} = \ket{n(t)} \rbrace}$ is any complete set of bases for the states in the Hilbert space, one can always find a metric $G(t)$ that satisfies
			\begin{align}
				\sum_n \ket{n(t)}\bra{n(t)} G(t) = \mathbbm{1}, \label{AChoice}
			\end{align}
			and vice versa. Note that the $\ket{n(t)}$ above are not limited to the eigenkets of the Hamiltonian. In fact, Eq.~\eqref{AChoice} is merely a direct generalization of the completeness relation:
			\begin{align}
				\sum_n \Ket{n(t)} \Bra{n(t)} = \mathbbm{1}. \label{CompletenessRelation}
			\end{align}	
				
			The $G(t)$ can be proven to be Hermitian and positive definite, and is also a solution of Eq.~\eqref{MetricEquation}. Using Eq.~\eqref{AChoice}, one can find a corresponding metric $G(t)$ using any complete set of bases. Interestingly, if the Hamiltonian eigenstates form a complete set of bases, the corresponding $G(t)$ in Eq.~\eqref{AChoice}, using the eigenstates as the set $\lbrace\Ket{n(t)}\rbrace$, is the same metric in BQM. Appendix \ref{PTSymmetricExample} provides some examples on how Eq.~\eqref{AChoice} can be used in finding $G(t)$.
		
	\section{Generalized Operators in NHQM \label{GeneralizedOperators}}
	
		As discussed above, in general, the generalized inner products are different from the conventional ones. As will be shown shortly, the corresponding operators also need to be modified. For example, the roles of unitary operators are no longer special since, in general, they do not leave the norm squared of states invariant. In the following we focus on the modifications of some common operators that will be proven useful for later use. Basic differences between conventional and non-Hermitian QM are summarized in Table~\ref{TableI}.
	
		\subsection{Adjoint Operators}
		
			In CQM, the ``bra'' vectors are defined to be the Hermitian conjugate of the corresponding ``ket'' vectors and, therefore, the adjoint operators are the conventional Hermitian conjugate of the original operators. In the modified Hilbert space, however, this property is different. Assuming that $\mathcal{O}(t)$ is an operator acting on the ket vector, we find that
			\begin{align}
				\begin{split}
					& \Braket{\psi(t)}{\mathcal{O}(t)\phi(t)} = \bra{\psi(t)} G(t) \mathcal{O}(t) \ket{\phi(t)}\\
					& \quad = \bra{\psi(t)} G(t) \mathcal{O}(t) G^{-1}(t) G(t) \ket{\phi(t)}\\
					& \quad = \bra{G^{-1}(t) \mathcal{O}^\dagger (t) G(t) \psi(t)} G(t) \ket{\phi(t)}\\
					& \quad = \Braket{G^{-1}(t) \mathcal{O}^\dagger (t) G(t) \psi(t)}{\phi(t)},
				\end{split}
			\end{align}
			which shows that the adjoint of $O(t)$ is 
			\begin{align}
				\mathcal{O}^\sharp (t)= G^{-1}(t) \mathcal{O}^\dagger (t) G(t), \label{AdjointCondition}
			\end{align}
			where $^\sharp$ stands for the corresponding adjoint operator in the modified Hilbert space and $^\dagger$ is the standard Hermitian conjugate. A quick observation shows that applying the Hermitian conjugate twice leads, again, to the original operator:
			\begin{align}
				\begin{split}
					& \left[ \mathcal{O}^\sharp (t) \right]^\sharp = G^{-1}(t) \left[ \mathcal{O}^\sharp (t) \right]^\dagger G(t)\\
					& \quad = G^{-1}(t) \left[ G^{-1}(t) \mathcal{O}^\dagger (t) G(t) \right]^\dagger G(t)\\
					& \quad = G^{-1}(t) G(t) \mathcal{O} (t) G^{-1}(t) G(t) = \mathcal{O}(t).
				\end{split}
			\end{align}
			The second last equality utilizes the Hermiticity of $G(t)$, i.e., ${G^\dagger (t) = G(t)}$.
			
			Because observables are among the most important ingredients in QM and widely-believed to be self-adjoint operators, it is worth finding the generalized ``Hermitian operators.'' By using Eq.~\eqref{AdjointCondition}, it is not hard to see that
			\begin{align}
				\mathcal{O}^\sharp (t) = \mathcal{O}(t) \Rightarrow ~ \mathcal{O}^\dagger (t) G(t) = G(t) \mathcal{O} (t). \label{HermitianOperators}
			\end{align}

			As shown in Appendix \ref{HermitianCase}, the metric operators can be set to unity for all Hermitian systems. In such cases, the conditions for adjoint operators and Hermitian operators in CQM can be recovered using Eqs.~\eqref{AdjointCondition} and \eqref{HermitianOperators}.
	
		\subsection{Generalized Unitary Operators}
		
			It is well known that a unitary transformation in CQM does not change the value of the inner products. To be more specific, let $U(t)$ be a unitary operator, then
			\begin{align}
				\begin{split}
					& \bra{U(t) \psi (t)}\ket{U(t) \phi (t)} = \braket{\psi(t)}{U^\dagger(t) U(t) \phi (t)}\\
					& \quad = \braket{\psi(t)}{U^{-1}(t) U(t) \phi (t)} = \bra{\psi (t)}\ket{\phi (t)}.
				\end{split}
			\end{align}

			However, in a Hilbert space where the metric ceases to be $\mathbbm{1}$, the unitarity loses its meaning and has to be modified. For the inner product to be invariant under a linear action $\mathcal{U}(t)$, the operator should satisfy 
			\begin{align}
				\begin{split}
					& \Braket{\psi(t)}{\phi (t)} = \Braket{\mathcal{U}(t) \psi (t)}{\mathcal{U}(t) \phi (t)}\\
					& \quad = \Braket{\psi (t)}{\mathcal{U}^{-1}(t) \mathcal{U}(t) \phi (t)}.
				\end{split}
			\end{align}
			Replacing Eq.~\eqref{AdjointCondition} with
			\begin{align}
				\begin{split}
					\mathcal{O} \rightarrow \mathcal{U}^{-1}(t),\\	
					\mathcal{O}^\sharp (t) \rightarrow \mathcal{U}(t),
				\end{split}
			\end{align}
			shows that the operators satisfying
			\begin{align}
				\mathcal{U}^{\dagger}(t) = G(t) \mathcal{U}^{-1}(t) G^{-1}(t)
			\end{align}
			leave the inner products invariant. Indeed, these are the operators that leave the metric invariant, i.e.,
			\begin{align}
				\mathcal{U}^\dagger (t) G(t) \mathcal{U}(t) = G(t). \label{UnitaryOperators}
			\end{align}
			
		\subsection{Generalized Density Matrices \label{DensityMatrices}}
		
			The modification of the density matrices should be quite obvious,
			\begin{align}
				\begin{split}
					& \rho_\text{CQM}(t) = \sum_i p_i \ket{\psi_i(t)} \bra{\psi_i(t)}\\
					& \rightarrow \rho(t) = \sum_i p_i \Ket{\psi_i(t)} \Bra{\psi_i(t)},
				\end{split}\label{DefDensityMatrices}
			\end{align}
			where $\rho_\text{CQM}(t)$ denotes a standard density matrix in CQM, while $\rho(t)$ is a generalized density matrix (GDM) in NHQM, and $p_i$ is the probability of obtaining the state $\Ket{\psi_i(t)}$. It can be proven that this operator indeed satisfies the properties of density matrices (i.e., self-adjoint, positive semi-definite, and trace can be set to unity)~\cite{BergouDensityMatrix} and leads to physically reasonable outcomes. All the detailed derivations and proofs are in Appendix \ref{AppDensityMatrices}.
			
			Note that since the trace of the conventional density matrix is not necessarily time independent, many studies introduce a normalized density operator, ${\displaystyle \rho_\text{N}(t) \equiv \rho_\text{CQM}(t) / \tr\left[ \rho_\text{CQM}(t) \right]}$, where the time evolution obeys
			\begin{align}
				\begin{split}
					i \hbar\partial_t \rho_\text{N}(t) & = H(t) \rho_\text{N}(t) - \rho_\text{N}(t) H^\dagger (t)\\
					& \quad + \tr \left\lbrace \rho_\text{N}(t) \left[ H^\dagger (t) - H(t) \right] \right\rbrace \rho_\text{N}(t).
				\end{split}
			\end{align}
			Unlike the normalized density matrix, the density matrix studied here is more natural in the sense that the trace is already time independent by construction. The reason behind this is because the time evolution of the GDM is governed by
			\begin{align}
				i \hbar \partial_t \rho(t) = [ H(t) , \rho (t) ],
			\end{align}
			even for $H(t) \neq H^\dagger (t)$. By using the cyclic property of the trace, it is obvious that
			\begin{align}
				\partial_t \tr \rho(t) = 0.
			\end{align}
			
			As will be shown in Sec.~\ref{No-Go}, unlike a normalized density matrix, which leads to the violation of various no-go theorems, the GDM preserves the no-go theorems. Moreover, since in general the time evolution of the generalized density matrix is different from the normalized one, the validity of this formulation can be verified experimentally.
		
		\begin{table}[h]
			\renewcommand*{\arraystretch}{1.6}
			\begin{tabular}{|>{\centering\arraybackslash}m{0.22\textwidth}| >{\centering\arraybackslash}m{0.24\textwidth}|}
				\hline
				\bf Conventional operator & \bf Generalized operator in NHQM\\
				\hline
				Hermitian conjugate & Adjoint\\
				$\displaystyle O^\dagger (t)$ \vspace{0.05cm} & $\displaystyle \mathcal{O}^\sharp = G^{-1}(t) \mathcal{O}^\dagger (t) G(t)$ \vspace{0.05cm}\\
				\hline
				Hermitian & Self-adjoint\\
				$\displaystyle O^\dagger (t) = O(t)$ \vspace{0.05cm} & $\displaystyle \mathcal{O}^\dagger (t) G(t) = G(t) \mathcal{O}(t)$ \vspace{0.05cm}\\
				\hline
				Unitary & Generalized unitary\\
				$\displaystyle U^\dagger (t) U(t) = \mathbbm{1}$ \vspace{0.05cm} & $\displaystyle \mathcal{U}^\dagger (t) G(t) \mathcal{U}(t) = G(t)$ \vspace{0.05cm}\\
				\hline
				Density matrix & Generalized density matrix\\
				$\displaystyle \rho = \sum_i p_i \ket{\psi_i(t)} \bra{\psi_i(t)}$  \vspace{0.05cm}& $\displaystyle \rho = \sum_i p_i \Ket{\psi_i(t)} \Bra{\psi_i(t)}$ \vspace{0.05cm}\\
				\hline
			\end{tabular}
			\caption{List of conventional operators (left) and their corresponding operators (right) in the modified Hilbert space. \label{Operators}}
		\end{table}
				
	\section{No-Go Theorems Revisited in NHQM \label{No-Go}}
	
		This work has been inspired by our former theoretical and experimental works on the no-cloning theorem and its applications in standard QM~\cite{Oezdemir2007, Bartkiewicz2009, Bartkiewicz2010, Lemr2012, Bartkiewicz2013, Bartkiewicz2017}. We now prove the validity of the no-cloning theorem and other related no-go theorems also in NHQM.
		
		As mentioned in Sec.~\ref{Introduction}, applying CQM on non-Hermitian quantum systems leads to the apparent violation of many no-go theorems. Nevertheless, the roles played by these no-go theorems in the quantum world are of importance and should be preserved when extended to NHQM. For a reason to be explained shortly, the relations between these no-go theorems are intertwined, violating any of them can lead to the failure of other no-go theorems. In addition, these violations contradict some of the well-established notions in quantum mechanics.
		
		\begin{table}[h]
			\renewcommand*{\arraystretch}{1.6}
			\begin{tabular}{|>{\centering\arraybackslash}m{0.19\textwidth}| m{0.275\textwidth}|}
				\hline
				\bf No-go theorem \vspace{0.05cm} & \bf Description \vspace{0.05cm}\\
				\hline
				No signaling \vspace{0.05cm} & Information cannot be transferred using an entangled state alone. \vspace{0.05cm}\\
				\hline
				No cloning \vspace{0.05cm} & Duplicating an arbitrary state is impossible. \vspace{0.05cm}\\
				\hline
				No deleting \vspace{0.05cm} & Erasing an arbitrary state is impossible. \vspace{0.05cm}\\
				\hline
				No perfect discrimination of nonorthogonal states \vspace{0.05cm} & Arbitrary non-orthogonal states cannot be distinguished perfectly. \vspace{0.05cm}\\
				\hline
				No increase of quantum entanglement by local operations \vspace{0.05cm} & Local operations cannot strengthen the entanglement between two parties.\vspace{0.05cm}\\
				\hline
				The invariance of entanglement under local \PT-symmetric unitaries \vspace{0.05cm} & Entanglement is invariant under local unitary transformations. \vspace{0.05cm}\\
				\hline
			\end{tabular}
			\caption{List of no-go theorems in quantum information and their brief descriptions. \label{NoGoTable}}
		\end{table}
		
		\subsection{No-go theorems in QM}
        	\subsubsection{No-cloning theorem}
		
				The no-cloning theorem of Ghirardi~\cite{GhirardiNoCloning}, Wootters and Zurek~\cite{WoottersNoCloning}, and Dieks~\cite{DieksNoCloning}, states that unknown \emph{pure} quantum states cannot be perfectly copied. The no-broadcasting theorem of Barnum \emph{et al.}~\cite{Barnum1996, Barnum2007} is a \emph{mixed}-state generalization of the no-cloning theorem, stating that unknown \emph{mixed} quantum states cannot be perfectly copied.

			\subsubsection{No-Deleting Theorem}
			
				The no-deleting theorem of Pati and Braunstein~\cite{PatiNoDeleting,Zurek2000} states that unknown \emph{pure} quantum states cannot be deleted, which clearly is a principle complimenting the no-cloning theorem. We note that a generalization of the no-deleting theorem to the case of mixed states has not been proven or even precisely formulated yet.
			
			\subsubsection{No-Signaling Theorem}
				
				The no-signaling theorem (also referred to as the no-communication theorem) states that the quantum entanglement between two spatially separated particles cannot provide superluminal (i.e., faster than the speed of light in vacuum) communication~\cite{Ghirardi1980,Ghirardi1983,PeresBook} (for an experimental test see~\cite{DeAngelis2007}). This implies that the shared entanglement \emph{alone} cannot be used to transmit any useful information.
				
				The no-signaling and no-cloning theorems are closely related (see, e.g., \cite{Gisin1998, Ghosh1999, Hardy1999, Bruss2000, Sekatski2015}). The no-signaling theorem implies bounds on quantum cloning. And perfect quantum cloning machines (QCMs) would allow arbitrary fast signaling by quantum entanglement using, e.g., Herbert's communicator using EPR states~\cite{Herbert1982}. Actually, the no-cloning theorem was first formulated~\cite{GhirardiNoCloning, WoottersNoCloning, DieksNoCloning} to demonstrate that Herbert's superluminal communicator cannot work since there are no perfect QCMs.
				
				As shown by Gisin~\cite{Gisin1998}, a tight bound on the fidelity of QCMs is compatible with the no-signaling constraint and is equal to the fidelity of the universal QCM of Buzek and Hillery~\cite{Buzek1996}. A whole class of 1-to-2 optimal quantum cloning machines of single qubits can be obtained from the no-signaling theorem~\cite{Ghosh1999, Hardy1999}. Various types of approximate QCMs have been experimentally applied for testing the security of quantum cryptographic systems (e.g., in~\cite{Bartkiewicz2013}) including the security of quantum money~\cite{Bartkiewicz2017}.
				
				The no-cloning and no-signaling theorems can be derived from the no-go principle concerning quantum-state discrimination~\cite{Barnett2009, Bae2015}, which says that nonorthogonal quantum states cannot be perfectly discriminated, as a consequence of the superposition principle. Indeed, two nonorthogonal states of a quantum system have a nonzero overlap, which implies that it is impossible to unambiguously determine which of the states has been achieved by the system. In particular, a tight bound on the measurement, which can discriminate two nonorthogonal states without error, can be obtained from the no-signaling principle~\cite{Barnett2002}.
				
				The no-signaling principle is closely related to quantum entanglement, which is among the main resources in quantum information and quantum technologies of the second generation~\cite{GeorgescuQuantumTech}. A proper measure of entanglement must satisfy some basic physical laws~\cite{HorodeckiEntanglement}, including  no increase of entanglement under any local operations and the invariance of entanglement under local unitary operations. Surprisingly, two works showed that these basic properties of good entanglement measures of standard quantum mechanics can be violated in the \PT-symmetric quantum theory. Specifically, Refs.~\cite{ChenIncreasingEntanglement,PatiViolateInvariance} \emph{apparently} showed, respectively, that entanglement under local operations can be increased and entanglement under local unitary operations is not invariant under local \PT-symmetric unitary operations.
		
				These no-go principles have profound implications in quantum theory for both its fundamental aspects (e.g., quantum causality) and applications including quantum metrology, quantum communication, and quantum cryptography. In particular, the violation of the no-cloning theorem would imply, in a trivial way, the violation of the Heisenberg's uncertainty principle in quantum mechanics. Moreover, basic dynamical rules of quantum physics can be derived from its static properties and the non-signaling theorem~\cite{Simon2001}.
				
				The following shows the validity of some no-go theorems in NHQM using the formalism provided in the previous sections. But it should be stressed that our generalized formalism reduces to that of Bender \emph{et al.}~\cite{BenderComplex, BenderCPT, BenderMakingSense} at least for $\cal{CPT}$-symmetric finite-dimensional systems. Thus, our proofs of the no-go principles are also valid for the latter systems.
		
		\subsection{No-Cloning Theorem in NHQM}
	
			Quantum copying is not allowed~\cite{WoottersNoCloning,DieksNoCloning} in CQM. This section shows that the no-cloning theorem continues to hold in a non-Hermitian system. Instead of proving that there are no unitary operators (the ``generalized unitary operators'') that can copy any arbitrary state to a blank state, we prove the theorem by contradiction. We assume that $\mathcal{C}(t)$ is a cloning operator such that
			\begin{align}
				\mathcal{C}(t) \Ket{\psi(t)} \otimes \Ket{\text{E} (t)} = e^{i \theta[\psi(t)]} \Ket{\psi(t)} \otimes \Ket{\psi (t)}
			\end{align}
			for any $\Ket{\psi (t)}$, where $\Ket{\text{E} (t)}$ is a blank state and ${\theta[\psi(t)] \in \mathbb{R}}$ is some phase generated by the cloning process. Because it is a unitary operator of a direct product state, it must also satisfy
			\begin{align}
				\mathcal{C}^{\dagger}(t) G(t) = G(t) \mathcal{C}^{-1}(t),
			\end{align}
			where $G(t) = G_1(t) \otimes G_2(t)$, in which $G_1(t)$ and $G_2(t)$ are the metrics for $\Ket{\psi (t)}$ and $\Ket{\text{E} (t)}$, respectively. To make the proof as general as possible, $G_1 (t)$ is allowed to be different from $G_2 (t)$.
	
			Then following almost the same argument of the conventional proof of the no-cloning theorem~\cite{WoottersNoCloning,DieksNoCloning}, the inner product between two states becomes
			\begin{align}
				\begin{split}
					& \Braket{\psi(t)}{\phi(t)} = \Braket{\psi(t), \text{E} (t)}{\phi(t), \text{E} (t)}\\
					& \quad = \Bra{\psi(t), \text{E} (t)} \mathcal{C}^{-1}(t) \mathcal{C}(t) \Ket{\phi(t), \text{E} (t)}\\
					& \quad = \bra{\psi(t), \text{E} (t)} G(t)\mathcal{C}^{-1}(t) \mathcal{C}(t) \ket{\phi(t), \text{E} (t)}\\
					& \quad = \bra{\psi(t), \text{E} (t)} \mathcal{C}^{\dagger}(t) G(t) \mathcal{C}(t) \ket{\phi(t), \text{E} (t)}\\
					& \quad = e^{i\lbrace \theta [\phi(t)] - \theta [\psi(t)] \rbrace} \bra{\psi(t), \text{E} (t)} G(t) \ket{\phi(t), \text{E} (t)}\\
					& \quad = e^{i\lbrace \theta [\phi(t)] - \theta [\psi(t)] \rbrace} \Braket{\psi(t)}{\phi(t)} \Braket{\psi(t)}{\phi(t)},
				\end{split}
			\end{align}
			where $\Ket{\psi(t), \text{E} (t)}$ denotes $\Ket{\psi(t)} \otimes \Ket{\text{E} (t)}$ and $\ket{\psi(t), \text{E} (t)}$ denotes $\ket{\psi(t)} \otimes \ket{\text{E} (t)}$. Therefore, as in CQM, $\Ket{\psi(t)}$ and $\Ket{\phi(t)}$ have to be parallel or orthogonal with each other at time $t$. To be more precise, because it holds
			\begin{align}
				\Braket{\psi(t_1)}{\phi(t_1)} = \Braket{\psi(t_2)}{\phi(t_2)}
			\end{align}
			for any $t_1$ and $t_2$, the statement is true at any given moment of time. In other words, if two states are not parallel or orthogonal to each other at any given time, they can never be. Therefore, even for $\mathcal{C}(t)$ being a time-dependent operator, it is impossible for it to exist without leading to the contradiction. This completes the proof of the no-cloning theorem for general Hamiltonians.
			
		\subsection{No-Deleting Theorem in NHQM}

			Not only is copying prohibited, shredding is also forbidden in QM~\cite{PatiNoDeleting}, neither by direct deleting nor through ancilla states. To show this, we assume there is a linear operator $\mathcal{D}(t)$ that deletes any duplicated quantum state.
	
			Let us consider an arbitrary state $\Ket{ \psi(t) }$ and its duplicate. If deleting the duplicated state directly is possible, then one can write
			\begin{align}
				\mathcal{D}(t) \Ket{\psi (t), \psi (t)} = \Ket{\psi (t), \text{E} (t)}, \label{Deleting}
			\end{align}
			where $\Ket{\text{E} (t)}$ is, again, some blank state independent of $\Ket{\psi (t)}$. Contracting both sides of Eq.~\eqref{Deleting} with $\Bra{\psi (t), \text{E} (t)}$ gives
			\begin{align}
				\begin{split}
					& \Bra{\psi (t), \text{E} (t)} \mathcal{D}(t) \Ket{\psi (t), \psi (t)}\\
					& \quad = \Braket{\psi (t)}{\psi (t)} \Braket{\text{E} (t)}{\text{E} (t)}.
				\end{split}\label{Contracted}
			\end{align}
			However, repeating the process by replacing $\Ket{\psi (t), \psi (t)}$ with $\Ket{a \psi (t), a \psi (t)}$, where $a$ is some constant, gives
			\begin{align}
				\begin{split}
					& \Bra{a \psi (t), E (t)} \mathcal{D}(t) \Ket{a \psi (t), a \psi (t)}\\
					& \quad = \Braket{a \psi (t)}{a \psi (t)} \Braket{\text{E} (t)}{\text{E} (t)}.
				\end{split}\label{Contracted2}
			\end{align}
			Using the linearity of both the inner product and $\mathcal{D}(t)$, Eq.~\eqref{Contracted2} becomes
			\begin{align}
				\begin{split}
					& a |a|^2  \Bra{\psi (t), \text{E} (t)} \mathcal{D}(t) \Ket{\psi (t), \psi (t)}\\
					& \quad = |a|^2 \Braket{\psi (t)}{\psi (t)} \Braket{\text{E} (t)}{\text{E} (t)}.
				\end{split}\label{Contracted3}
			\end{align}
			By comparing Eq.~\eqref{Contracted3} with Eq.~\eqref{Contracted}, one finds that only when ${a = 0}$, ${a = 1}$, or ${\Braket{\text{E} (t)}{\text{E} (t)} = 0}$ gives a sensible solution. However, since $a$ is general, the only possible solution left is ${\Braket{\text{E} (t)}{\text{E} (t)} = 0}$. This leads to ${\Ket{\text{E}(t)}=0}$ due to the positive definiteness of the metric and implies
			\begin{align}
				\mathcal{D}(t) \Ket{\psi (t), \psi (t)} = 0.
			\end{align}	
			Hence, a direct deleting process does not exist.
	
			Another possible deleting process is through the aid of some ancilla state $\Ket{\text{A} (t)}$, which reads
			\begin{align}
				\mathcal{D}(t) \Ket{\psi (t), \psi (t), \text{A} (t)} = \Ket{\psi (t), \text{E} (t), \text{A}_{\Ket{\psi}} (t)},\label{DeletingAncilla}
			\end{align}
			where the subscript $\Ket{\psi}$ in $\Ket{\text{A}_{\Ket{\psi}} (t)}$ means that the state is an implicit function of $\Ket{\psi (t)}$. Contracting Eq.~\eqref{DeletingAncilla} with $\Bra{\psi (t), \text{E} (t), \text{A}_{\Ket{\psi}} (t)}$ gives
			\begin{align}
				\begin{split}
					& \Bra{\psi (t), \text{E} (t), \text{A}_{\Ket{\psi}} (t)}\mathcal{D}(t) \Ket{\psi (t), \psi (t), \text{A} (t) \vphantom{\text{A}_{a\Ket{\psi}}}}\\
					& \quad = \Braket{\psi (t), \text{E} (t), \text{A}_{\Ket{\psi}} (t)}{\psi (t), \text{E} (t), \text{A}_{\Ket{\psi}} (t)}.\label{DeletingAncilla1}
				\end{split}
			\end{align}
			By rescaling $\Ket{\psi(t)}$ to $\Ket{a\psi(t)}$ and repeating the procedure above results in
			\begin{align}
				\begin{split}
					& a |a|^2 \Bra{\psi (t), \text{E} (t), \text{A}_{a\Ket{\psi}} (t)}\mathcal{D}(t) \Ket{\psi (t), \psi (t), \text{A} (t) \vphantom{\text{A}_{a\Ket{\psi}}}}\\
					& \quad = |a|^2 \Braket{\psi (t), \text{E} (t), \text{A}_{a\Ket{\psi}} (t)}{\psi (t), \text{E} (t), \text{A}_{a\Ket{\psi}} (t)}.\label{DeletingAncilla2}
				\end{split}
			\end{align}
			
			Comparing Eqs.~\eqref{DeletingAncilla1} and \eqref{DeletingAncilla2} without $a$ being limited to 0 or 1, the only possible solutions left are ${\Ket{\text{E}(t)}=0}$, ${\Ket{\text{A}_{a\Ket{\psi}} (t)}=0}$, or ${\Ket{\text{A}_{a\Ket{\psi}} (t)} = a \Ket{\text{A}_{\Ket{\psi}} (t)}}$. The first two cases lead to the unwanted result ${\mathcal{D}(t) \Ket{\psi (t), \psi (t), \text{A} (t)} = 0}$. The third solution shows that $\Ket{\text{A}_{\Ket{\psi}} (t)}$ is linear in $\Ket{\psi (t)}$. This means that $\mathcal{D}(t)$ moves the duplicated $\Ket{\psi (t)}$ to the ancilla state, which is not a deleting operator but a quantum swapping operator up to a linear scrambling.
			
			In conclusion, neither direct deleting nor deleting through an ancilla state operator can exist. Therefore, a duplicated quantum state cannot be deleted through a linear operator in the NHQM.
		
		\subsection{No-Signaling Theorem in NHQM}
		
			To show that any local measurement performed on a quantum system, say A, renders no statistical effect on another quantum system, say B, in a general quantum system (no superluminal communication~\cite{GhirardiNoCloning}), the measurement operators have to be modified according to Table \ref{Operators}. Therefore, the measurements $M_{\text{A}j}(t)$ on system A satisfying the identity relation,
			\begin{align}
				\sum_j M_{\text{A}j}^\dagger (t) M_{\text{A}j} (t) = \mathbbm{1} ,
			\end{align}
			in the CQM have to be modified to
			\begin{align}
				& \sum_j \mathcal{M}_{\text{A}j}^\sharp (t) \mathcal{M}_{\text{A}j} (t) = \mathbbm{1}\\
				& \quad \Rightarrow \sum_j \mathcal{M}_{\text{A}j}^\dagger (t) G_\text{A} (t) \mathcal{M}_{\text{A}j} (t) = G_\text{A} (t),\label{LOCC}
			\end{align}
			for the measurements $M_{\text{A}j}(t)$ in NHQM.
			
			Let us consider a state
			\begin{align}
				\rho(t) = \sum_i p_i \Ket{\psi_i(t)}\Bra{\psi_i(t)}
			\end{align}
			and assume a measurement on system A, ${\widetilde{\mathcal{M}}_j (t) \equiv \mathcal{M}_{\text{A}j} (t) \otimes \mathbbm{1}_\text{B}}$. Then the reduced density matrix for system B becomes 
			\begin{align}
				\begin{split}
					& \rho'_\text{B} (t)= \tr_\text{A} \sum_{i,j} p_i \Ket{ \widetilde{\mathcal{M}}_j (t) \psi_i (t)} \Bra{\widetilde{\mathcal{M}}_j (t) \psi_i (t)}\\
					& \quad = \tr_\text{A} \sum _{i,j} p_i \ket{ \widetilde{\mathcal{M}}_j (t) \psi_i (t)} \bra{\widetilde{\mathcal{M}}_j (t) \psi_i (t)} G (t)\\
					& \quad= \tr_\text{A} \sum _{i,j} p_i \widetilde{\mathcal{M}}_j (t) \ket{\psi_i (t)} \bra{ \psi_i (t)} \widetilde{\mathcal{M}}_j^\dagger (t) G (t)\\
					& \quad = \tr_\text{A} \sum _{i,j} p_i \ket{\psi_i (t)} \bra{ \psi_i (t)} \widetilde{\mathcal{M}}_j^\dagger (t) G (t) \widetilde{\mathcal{M}}_j (t)\\
					& \quad = \tr_\text{A} \sum _{i} p_i \ket{\psi_i (t)} \bra{ \psi_i (t)} G (t)\\
					& \quad = \tr_\text{A} \sum _{i} p_i \Ket{\psi_i (t)} \Bra{ \psi_i (t)} = \rho_\text{B} (t),
				\end{split}\label{NoSignaling}
			\end{align}
			where ${G (t) = G_\text{A} (t) \otimes G_\text{B} (t)}$, with ${G_\text{A/B} (t)}$ being the metric for system A/B. The fourth equality in Eq.~\eqref{NoSignaling} uses the cyclic property of the trace and the fifth equality utilizes Eq.~\eqref{LOCC}, which renders
			\begin{align}
				\begin{split}
					& \widetilde{\mathcal{M}}_j^\dagger (t) G (t) \widetilde{\mathcal{M}}_j (t)\\
					& \quad = \left[\mathcal{M}_{\text{A}j}^\dagger (t) \otimes \mathbbm{1}_\text{B}\right] \left[ G_\text{A} (t) \otimes G_\text{B} (t) \right] \left[\mathcal{M}_{\text{A}j} (t) \otimes \mathbbm{1}_\text{B}\right]\\
					& \quad = \left[ G_\text{A} (t) \otimes G_\text{B} (t) \right] = G (t).
				\end{split}
			\end{align}	
			
			Equation \eqref{NoSignaling} shows that the reduced density matrix ${\rho'_\text{B} (t)}$ in system B is not affected by the local measurements performed on system A.
		
		\subsection{Entanglement Invariance Under Local Unitary Transformation in NHQM}
		
			One of the most exotic phenomena in QM, compared to classical ones, is entanglement. A good entanglement measure is a function quantifying the entanglement between two systems, which satisfies a number of conditions~\cite{HorodeckiEntanglement}. In particular, it should be invariant under local unitary transformations~\cite{VidalInvairant}. We discuss here the entanglement of formation~\cite{BennettEntanglementOfFormation}, which is closely related to the Wootters concurrence~\cite{WootersConcurrence}, between systems A and B:
			\begin{align}
				E \left[ \rho (t) \right] = \inf_{\lbrace p_i, \Ket{\psi_i} \rbrace \circeq \rho} \sum_i p_i E_\text{P} \left[ \Ket{\psi_i (t)} \right],\label{FormationOfEntanglement}
			\end{align}
			where $\lbrace p_i , \Ket{\psi_i} \rbrace \circeq \rho$ means a set of probabilities and pure states such that $\displaystyle \rho (t) = \sum_i p_i \Ket{\psi_i (t)} \Bra{\psi_i (t)}$, and $E_\text{P}$ is the entropy of entanglement for pure states defined as
			\begin{align}
				\begin{split}
					& E_\text{P} \left[ \Ket{\psi(t)} \right] = - \tr \left[ \rho_\text{A}  (t) \ln  \rho_\text{A} (t) \right]\\
					& \quad =  - \tr \left[ \rho_\text{B}  (t) \ln  \rho_\text{B} (t) \right],
				\end{split}\label{EntropyOfEntengelment}
			\end{align}
			where $\rho_\text{A/B} (t) = \tr_\text{B/A} \Ket{\psi(t)} \Bra{\psi(t)}$. The second equality in Eq.~\eqref{EntropyOfEntengelment} originates from the self-adjointness and semipositiveness of the GDM, which renders the eigenvalues to be real and non-negative.
			
			By the symmetry of the entanglement of formation [Eq.~\eqref{EntropyOfEntengelment}], without loss of generality, a unitary transformation $\mathcal{U}_\text{A} (t)$ is assumed to act on system A that leads to
			\begin{align}
				\begin{split}
					& \rho^{(\mathcal{U})}_\text{A} (t) = \tr_\text{B} \left[ \Ket{\mathcal{U} (t) \psi(t)} \Bra{\mathcal{U} (t) \psi(t)} \right]\\
					& \quad = \tr_\text{B} \left[ \mathcal{U} (t) \Ket{\psi(t)} \Bra{\psi(t)}\mathcal{U}^{-1} (t) \right]\\
					& \quad = \mathcal{U} (t) \tr_\text{B} \left[  \Ket{\psi(t)} \Bra{\psi(t)} \right] \mathcal{U}^{-1} (t)\\
					& \quad = \mathcal{U}_\text{A} (t) \rho_\text{A} (t) \mathcal{U}^{-1}_\text{A} (t),
				\end{split}\label{ReducedDensityMatirxA2}
			\end{align}
			where $\mathcal{U} (t) = \mathcal{U}_\text{A} (t) \otimes \mathbbm{1}_\text{B}$. Using Eq.~\eqref{ReducedDensityMatirxA2}, the entropy of entanglement becomes
			\begin{align}
				\begin{split}
					& E_\text{P} \left[ \Ket{\mathcal{U} (t) \psi(t)} \right] = - \tr \left[ \rho^{(\mathcal{U})}_\text{A} (t) \ln \rho^{(\mathcal{U})}_\text{A} (t) \right]\\
					& \quad = - \tr \left\lbrace \mathcal{U}_\text{A} (t) \rho_\text{A} (t) \mathcal{U}^{-1}_\text{A} (t) \ln \left[ \mathcal{U}_\text{A} (t) \rho_\text{A} (t) \mathcal{U}^{-1}_\text{A} (t) \right] \right\rbrace\\
					& \quad = - \tr \left\lbrace \mathcal{U}_\text{A} (t) \rho_\text{A} (t) \ln \left[ \rho_\text{A} (t) \right] \mathcal{U}^{-1}_\text{A} (t) \right\rbrace\\
					& \quad = - \tr \left\lbrace \rho_\text{A} (t) \ln \left[ \rho_\text{A} (t) \right] \right\rbrace = E_\text{P} \left[ \Ket{ \psi(t)} \right].
				\end{split}\label{EntropyOfEntanglementA}
			\end{align}
			A direct consequence of Eq.~\eqref{EntropyOfEntanglementA} is that
			\begin{align}
				\begin{split}
					& E \left[ \mathcal{U} (t) \rho (t) \mathcal{U}^{-1} (t) \right]\\
					& \quad  \inf_{\lbrace p_i, \Ket{\mathcal{U} \psi_i} \rbrace \circeq \mathcal{U} \rho \mathcal{U}^{-1}} \sum_i p_i E_\text{P} \left[ \Ket{\mathcal{U} (t) \psi_i (t)} \right]\\
					& \quad = \inf_{\lbrace p_i, \Ket{\psi_i (t)} \rbrace \circeq \rho} \sum_i p_i E_\text{P} \left[ \Ket{\mathcal{U} (t) \psi_i (t)} \right]\\
					& \quad = \inf_{\lbrace p_i, \Ket{\psi_i (t)} \rbrace \circeq \rho} \sum_i p_i E_\text{P} \left[ \Ket{\psi_i (t)} \right] = E \left[ \rho (t) \right],
				\end{split}\label{LocalUnitaryInvariance}
			\end{align}
			where $\mathcal{U}$ can be either ${\mathcal{U}_\text{A} \otimes \mathbbm{1}_\text{B}}$ or ${\mathbbm{1}_\text{A} \otimes \mathcal{U}_\text{B}}$. In other words, Eq.~\eqref{LocalUnitaryInvariance} confirms that the entanglement of formation is invariant under a local unitary transformation also in NHQM.
			
		\subsection{No Entanglement Increasing Under Local Operations in NHQM}
		
			It is known that it is impossible to increase the entanglement between two systems by performing any local operation~\cite{BennettNoIncrease}. This can be shown in NHQM by considering the entanglement of formation of an arbitrary density matrix in Eq.~\eqref{FormationOfEntanglement}. Assuming $\left \lbrace \widetilde{p}_i , \Ket{\widetilde{\psi}_i (t)} \right\rbrace$ is the set such that 
			\begin{align}
				\rho (t) = \sum_i \widetilde{p}_i \Ket{\widetilde{\psi}_i (t)} \Bra{\widetilde{\psi}_i (t)}
			\end{align}
			and
			\begin{align}
				E \left[ \rho (t) \right] = \sum_i \widetilde{p}_i E_\text{P} \left[\Ket{\widetilde{\psi}_i(t)} \right],
			\end{align}
			after performing local operations, the density matrix $\rho(t)$ becomes
			\begin{align}
				\rho' (t) = \sum_j \mathcal{M}_j \rho (t) G^{-1} (t) \mathcal{M}^\dagger_j G(t).
			\end{align}
			Without loss of generality, assuming that measurements $\mathcal{M}_j$ are only performed on the system B, i.e., $\mathcal{M}_j = \mathbbm{1}_A \otimes M_{Bj}$. Thus,
			\begin{align}
				\Ket{[ij](t)} = \frac{1}{\sqrt{p_{[ij]}}}\Ket{\mathcal{M}_j \widetilde{\psi}_i (t)},
			\end{align}
			where 
			\begin{align}
				p_{[ij]} = \Braket{\mathcal{M}_j \widetilde{\psi}_i (t)}{\mathcal{M}_j \widetilde{\psi}_i (t)},
			\end{align}
			is still a pure state. In other words, $\left\lbrace \widetilde{p}_i p_{[ij]} , \Ket{[ij] (t)} \right\rbrace$ is a set of pure states such that
			\begin{align}
				\rho'(t) = \sum_{ij} \widetilde{p}_i p_{[ij]} \Ket{[ij] (t)}\Bra{[ij] (t)}.
			\end{align}
			The entropy of this density matrix is
			\begin{align}
				\begin{split}
					E(\rho') = & \inf_{\lbrace p'_i, \Ket{\psi'_i} \rbrace\circeq \rho'} \sum_i p'_i E_\text{P} \left[\Ket{\psi'_i(t)}\right]\\
					\leq & \sum_{i,j} \widetilde{p}_i p_{[ij]} E_\text{P} \left( \Ket{[ij] (t)} \right]\\
					\leq & \sum_{i} \widetilde{p}_i E_\text{P} \left[ \sum_j \sqrt{p_{[ij]}} \Ket{[ij] (t)}\right]\\
					= & \sum_i \widetilde{p}_i E_\text{P} \left[ \Ket{\widetilde{\psi}_i (t)} \right] = E \left[ \rho (t) \right],
				\end{split} \label{NoIncrease}
			\end{align}
			where the first inequality comes from the definition of infimum and the second comes from the concavity of the von Neumann entropy. The other relations in Eq.~\eqref{NoIncrease} result from
			\begin{align}
				\begin{split}
					& \tr_B \sum_j p_{[ij]} \Ket{[ij] (t)} \Bra{[ij] (t)}\\
					& \quad = \tr_B \sum_j \Ket{\mathcal{M}_j \widetilde{\psi}_i (t)}\Bra{\mathcal{M}_j \widetilde{\psi}_i (t)}\\
					& \quad = \tr_B \sum_j\mathcal{M}_j \Ket{\widetilde{\psi}_i (t)}\Bra{\widetilde{\psi}_i (t)} G^{-1} (t) \mathcal{M}^\dagger_j G(t)\\
					& \quad = \tr_B \sum_j\Ket{\widetilde{\psi}_i (t)}\Bra{\widetilde{\psi}_i (t)} G^{-1} (t) \mathcal{M}^\dagger_j G(t) \mathcal{M}_j\\
					& \quad = \tr_B \Ket{\widetilde{\psi}_i (t)}\Bra{\widetilde{\psi}_i (t)},
				\end{split}
			\end{align}
			so that
			\begin{align}
				\begin{split}
					& E_\text{P} \left[ \sum_j \sqrt{p_{[ij]}} \Ket{[ij] (t)}\right]\\
					& \quad = - \tr \bigg\lbrace \Big[ \tr_B \sum_j p_{[ij]} \Ket{[ij] (t)} \Bra{[ij] (t)} \Big]\\
					& \quad \quad ~ \cdot \ln \Big[ \tr_B \sum_j p_{[ij]} \Ket{[ij] (t)} \Bra{[ij] (t)}\Big] \bigg\rbrace\\
					& \quad = - \tr \bigg\lbrace \Big[ \tr_B \Ket{\widetilde{\psi}_i (t)}\Bra{\widetilde{\psi}_i (t)} \Big]\\
					& \quad \quad ~ \cdot \ln \Big[ \tr_B \Ket{\widetilde{\psi}_i (t)}\Bra{\widetilde{\psi}_i (t)} \Big] \bigg\rbrace\\
					& \quad = E_\text{P} \left[ \Ket{\widetilde{\psi}_i (t)} \right].
				\end{split}
			\end{align}
	
			Therefore, inequalities in Eq.~\eqref{NoIncrease} show that the entropy of formation under local operations cannot be greater than the original one in NHQM.
			
	\section{Conclusion}
	
		The notion of probability in finite-dimensional QM is closely related to the corresponding inner product. The total probability of a system should add up to unity. Any changes to the total probability of a system do not make any sense. The generalized or metricized inner product in finite-dimensional NHQM preserves the idea of the probability conservation by borrowing the concepts from fiber bundles. In this paper, the representation of quantum states are taken to be vectors as usual. Dual states (i.e., covectors), on the other hand, besides taking Hermitian conjugation, are corrected by a ``metric operator'' so that they become $\bra{\psi(t)} G(t)$. A few instructive examples, showing how to effectively calculate the metric, are given in Appendix \ref{Examples}.
		
		With this metric, some original definition of the operators lose their physical meaning. Hence, many studies that claimed that some of the no-go theorems are violated in non-Hermitian quantum systems might be a false alarm due to the use of the conventional definition. The generalized definitions of many related operators are discussed and derived in this paper. Note that Refs.~\cite{ZnojilPTFoundamental, BrodyConsistency} include the proofs of only the no-signaling theorem, or more precisely, the proofs that the apparent violation of the no-signaling theorem described in \cite{LeeYesSignaling} does not hold. We emphasize that our proof of the no-signaling theorem is simpler and more general, because it is valid even for non-\PT-symmetric non-Hermitian finite-dimensional systems. More importantly, we have also proved the other five no-go theorems, as listed in Table \ref{NoGoTable}, which are of fundamental importance for quantum information processing.
		
		We stress that our paper is devoted to the no-go theorems in the fully quantum regime of NHQM. Otherwise, in the semiclassical and classical regimes of NHQM, there is, e.g., no quantum entanglement (no-signaling, no increase of quantum entanglement by local operations, and the invariance of entanglement under local \PT-symmetric unitaries) and, moreover, perfect cloning, perfect deleting, and perfect discrimination of non-orthogonal classical states are allowed. Although how far this formalism can reach is an open question, with these definitions, many of the no-go theorems in quantum information, including no-cloning, no-deleting, along with a few others, are proven to be valid in finite-dimensional NHQM systems, including the well-known $\cal{CPT}$ systems, pseudo-Hermitian systems, and those where BQM is valid. Therefore, NHQM can still be a candidate for a fundamental physics theory.

	\section*{ACKNOWLEDGEMENT}
		We thank Prof. Dorje Brody and Prof. Max Lein for their useful comments on the first version of our manuscript. C.Y.J. and G.Y.C. are supported partially by the National Center for Theoretical Sciences and Ministry of Science and Technology, Taiwan, through Grant No. MOST 108-2112-M-005-005. F.N. is supported in part by the: MURI Center for Dynamic Magneto-Optics via the Air Force Office of Scientific Research (AFOSR) (Grant No. FA9550-14-1-0040), Army Research Office (ARO) (Grant No. W911NF-18-1-0358), Asian Office of Aerospace Research and Development (AOARD) (Grant No. FA2386-18-1-4045), Japan Science and Technology Agency (JST) (Q-LEAP program and CREST Grant No. JPMJCR1676), Japan Society for the Promotion of Science (JSPS) (JSPS-RFBR Grant No. 17-52-50023, and JSPS-FWO Grant No. VS.059.18N), RIKEN-AIST Challenge Research Fund, and the Foundational Questions Institute (FQXi), and the NTT PHI Laboratory.
		
	\appendix
	
		\section{Appnedix: Examples \label{Examples}}

		For pedagogical purposes we give in this Appendix a few simple but important examples. The readers are also referred to various instructive examples presented in the existing correct formulations of the NHQM theory. These include the
mathematics-oriented Ref.~\cite{Bagarello2015} and the more physics-oriented Ref.~\cite{Mostafazadeh2010}, as well as a recent monograph \cite{Bender2019}.
		
		This part begins with a trivial check by showing that the conventional inner product can be reproduced using the metricized inner product. Then a seemingly trivial non-Hermitian case is discussed. The solution to this case might be trivial; however, it shows some important physical insights of this inner product. Last but not least, some methods for finding the metric for a simple \PT-symmetric Hamiltonian are demonstrated. The Hamiltonian is easy to work with and shows all different aspects of the inner product.
		
		\subsection{Hermitian Hamiltonian \label{HermitianCase}}
		
			It is not hard to see that the conventional inner product for NHQM is a special case of the inner product with $G(t) = \mathbbm{1}$. To see if $G(t)=\mathbbm{1}$ is a proper metric, it must satisfy all the requirements stated in Sec.~\ref{InnerProductScetion}. First, $\mathbbm{1}$ is obviously Hermitian and positive definite. Second, $G(t) = \mathbbm{1}$ is indeed a trivial solution to Eq.~\eqref{MetricEquation} since $H=H^\dagger$. This means that the NHQM inner product of a Hermitian Hamiltonian can be
			\begin{align}
				\Braket{\psi_1 (t)}{\psi_2 (t)} = \bra{\psi_1 (t)} \mathbbm{1}\ket{\psi_2 (t)} = \bra{\psi_1 (t)}\ket{\psi_2 (t)}.
			\end{align}
			It is not surprising that, the inner product for CQM is then recovered.
		\subsection{A Trivial Non-Hermitian Hamiltonian\label{TrivialNonHermitian}}
			
			The next example is a NHQM Hamiltonian of a one-dimensional quantum space, namely
			\begin{align}
				H = \omega - i \frac{\Gamma}{2}.
			\end{align}
			The most general Hermitian and positive-definite solution to Eq.~\eqref{MetricEquation} in this case is
			\begin{align}
				G(t) = G_0 \exp(t \Gamma) \label{UnboundedG}
			\end{align}
			with $G_0 > 0$. Although this example seems mathematically trivial, it is actually non-trivial for further understanding the underlying physics concept. For example, Eq.~\eqref{UnboundedG} is an unbounded operator which is usually discarded due to physical constraints (i.e., no infinities allowed) and, therefore, there is no solution to Eq.~\eqref{MetricEquation} with this Hamiltonian (or at least the domain of $t$ has to be carefully chosen so that the solution is bounded in the region). Furthermore, the most general state for this Hamiltonian is
			\begin{align}
				\Ket{\psi (t)} = \psi_0 \exp \left(-i t \omega - t \frac{\Gamma}{2} \right),
			\end{align}
			which also seems to suffer from the unboundedness. However, neither $G(t)$ nor $\Ket{\psi(t)}$ alone is physical, so that $G(t)$ and $\Ket{\psi(t)}$ are not necessarily bounded. But when it comes to the inner product, not only it is bounded but remains a constant in time, namely
			\begin{align}
				\begin{split}
					& \Braket{\psi (t)}{\psi (t)} = \left( \overline{\psi_0} e^{i t \omega - t \Gamma/2} \right) \left( G_0 e^{t \Gamma} \right) \left( \psi_0 e^{-i t \omega - t \Gamma/2} \right)\\
					& \quad = |\psi_0|^2 G_0 = \Braket{\psi (0)}{\psi (0)}.
				\end{split}
			\end{align}
			Consequently, once the norm squared is normalized at some time, it stays normalized at any time. One might be expecting the ``probability'' to decay, but there is only one state that can be measured in the system; hence, the probability of measuring that state at every instant should be 100\%. To find this ``decay,'' the state has to be compared with the ``past'' state.
			
		\subsection{\PT-Symmetric Hamiltonian\label{PTSymmetricExample}}
			
			Our example is a two-dimensional quantum space with an interacting \PT-symmetric Hamiltonian. This example is simple but conceptually rich and interesting since it shows almost all kinds of time dependences of the metric (from constant, oscillating, polynomial, to exponential growth or decay). A Hamiltonian of this kind was first given in~\cite{BenderComplex}
			\begin{align}
				H = \begin{pmatrix}
					r e^{i \theta} & s\\
					s & r e^{-i \theta}
				\end{pmatrix}, \label{Bender2DPTHamiltonian}
			\end{align}
			with $r,~\theta ,~s \in \mathbb{R}$ and $s \neq 0$.
			
			A quick inspection of this Hamiltonian shows that it becomes nondiagonalizable when $s^2 = r^2 \sin \theta$, and, as will be discussed shortly, only when $s^2 > r^2 \sin^2 \theta$ does it allow a constant metric. Therefore, the metric has to be separately discussed in three different situations, namely, $s^2 > r^2 \sin^2 \theta$, $s^2 < r^2 \sin^2 \theta$, and $s^2 = r^2 \sin^2 \theta$, which correspond to the \PT-unbroken region, \PT-broken region, and an exceptional point, respectively. Note that the Hamiltonian is continuous in all the parameters.
			
			There are many ways to find the metric,
			\begin{align}
				G(t) = \begin{pmatrix}
						g_{11} (t) & g_{12} (t)\\
						g_{21} (t) & g_{22} (t)
					\end{pmatrix};
			\end{align}
			for comparison, two methods are used in this Appendix: a brute-force method and a method based on the generalized-completeness-relation method, given by Eq.~\eqref{AChoice}. The brute-force method corresponds to solving Eq.~\eqref{MetricEquation} directly and finding the relations between the undetermined coefficients. This method gives the most general metric, but it is sometimes really hard to effectively apply it. The generalized-completeness-relation method is slightly less general, but it shows some important physical features of the metric.
			
			\subsubsection{\PT Unbroken Region}
			
				We first deal with the case for $s^2 > r^2 \sin^2 \theta$. It will be shown shortly that the Hamiltonian in Eq.~\eqref{Bender2DPTHamiltonian} indeed allows a constant metric in the region. As the first non-trivial example, our discussion will be in slightly more detail.\\~
				
				\subsubsubsection{Brute-Force Method:}
					The general solutions to Eq.~\eqref{MetricEquation} with the Hamiltonian given in Eq.~\eqref{Bender2DPTHamiltonian} is
					\begin{align}
						\begin{split}
							& g_{11} (t) = -\mathcal{C}_ + e^{i \phi_+(t)} + \mathcal{C}_-e^{-i \phi_+(t)} + \mathcal{C}_1,\\
							& g_{22} (t) = \mathcal{C}_+ e^{i \phi_-(t)} - \mathcal{C}_-e^{-i \phi_-(t)} + \mathcal{C}_1,\\
							& g_{12} (t) = \mathcal{C}_2 - \mathcal{C}_+ e^{i \phi_0(t)} - \mathcal{C}_- e^{-i \phi_0(t)} - i \mathcal{C}_1 \sin \alpha,\\
							& g_{21} (t) = 2 \mathcal{C}_2 - g_{12}(t) ,
						\end{split}
					\end{align}
					where
					\begin{align}
						& \sin \alpha \equiv \dfrac{r}{s} \sin \theta,\\
						&\phi_\pm(t) \equiv \phi_0(t) \pm \alpha,\\
						& \phi_0(t) \equiv 2 t s \cos \alpha,
					\end{align}
					and ${\mathcal{C}_{\pm,1,2} \in \mathbb{C}}$ are undetermined constants. The Hermiticity of the metric simplifies the components to
					\begin{align}
						\begin{split}
							& g_{11} (t) = - A \cos \phi_+(t) + B \sin \phi_+(t) + C,\\
							& g_{22} (t) = A \cos \phi_-(t) - B \sin \phi_-(t) + C,\\
							& g_{12} (t) = D - i \left[ A \sin \phi_0(t) + B \cos \phi_0(t) + C \sin \alpha \right],\\
							& g_{21} (t) = g_{12}^*(t),
						\end{split}
					\end{align}
					where $A,~B,~C,~D \in \mathbb{R}$. Finally, it can be shown that the positivity of the metric requires ${C > \sqrt{A^2 + B^2}}$ and ${\left(C^2 - A^2 - B^2 \right) \cos^2 \alpha > D^2}$.
				
					One can easily find that, when ${A = B = 0}$, ${G(t)}$ becomes a constant metric. Furthermore, Bender's inner product can be reproduced by setting ${A,~B,~D}$ to zero and ${C=1/\cos \alpha}$, which results in
					\begin{align}
						G(t) = \frac{1}{\cos \alpha} \begin{pmatrix}
							1 & - i \sin \alpha\\
							i \sin \alpha & 1
						\end{pmatrix}.
					\end{align}
					
					\subsubsubsection{Method Based on the Generalized Completeness Relation:}
					
						The completeness relation in CQM is $\displaystyle {\sum_n \ket{n}\bra{n} = \mathbbm{1}}$, where the kets $\ket{n}$ form a complete set of bases. But in a non-Hermitian system, this relation has to be generalized to 
						\begin{align}
							& \sum_n \Ket{n(t)}\Bra{n(t)} = \mathbbm{1}\\
							\Rightarrow & \sum_n \ket{n(t)}\bra{n(t)} G(t) = \mathbbm{1}.
						\end{align}
						Note that the time dependence is shown explicitly in the generalized completeness relation. In fact, the time dependence cancels out directly in the conventional relation and so does the non-Hermitian case. This relation allows one to find a metric including its time dependence. Two different choices of complete sets of bases are worked out in the following.
					
							\paragraph{(1) Standard Choice}~\\
						
								Since the Hamiltonian in Eq.~\eqref{Bender2DPTHamiltonian} is diagonalizable in this region, the eigenvectors form a complete basis set. These eigenvectors can be used as the bases at $t=0$,
								\begin{align}
									\ket{\mathbf{1}(0)} = a\begin{pmatrix}
										e^{i\alpha / 2}\\
										e^{-i\alpha / 2}
									\end{pmatrix} ~ \text{and} ~ \ket{\mathbf{2}(0)} = b\begin{pmatrix}
										e^{-i\alpha / 2}\\
										-e^{i\alpha / 2}
									\end{pmatrix},
								\end{align}
								where $\cos \alpha = \sqrt{1 - \left(\frac{r}{s}\right)^2 \sin^2 \theta}$. The time evolutions of these vectors are
								\begin{align}
									\begin{split}
										\begin{array}{c}
											\displaystyle \ket{\mathbf{1}(t)} = a~ \exp(- i \lambda_+ t ) \begin{pmatrix}
												e^{ i\alpha / 2 }\\
												e^{-i \alpha / 2}
											\end{pmatrix},\\
											\displaystyle \ket{\mathbf{2}(t)} = b~ \exp(- i \lambda_- t )\begin{pmatrix}
												e^{-i \alpha / 2}\\
												- e^{i \alpha / 2}
											\end{pmatrix},
										\end{array}
									\end{split}
								\end{align}
								where $\lambda_\pm = \sqrt{r^2 - s^2 \sin^2 \alpha}\; \pm s \cos \alpha$. The generalized completeness relation becomes
								\begin{align}
									\begin{split}
										& \mathbbm{1} = \left[ \ket{\mathbf{1}(t)} \bra{\mathbf{1}(t)} + \ket{\mathbf{2}(t)} \bra{\mathbf{2}(t)}\right] G(t)\\
										& \Rightarrow G(t) = \left[ \ket{\mathbf{1}(t)} \bra{\mathbf{1}(t)} + \ket{\mathbf{2}(t)} \bra{\mathbf{2}(t)}\right]^{-1},
									\end{split}
								\end{align}
								where
								\begin{align}
									G(t) = \begin{pmatrix}
										\mathcal{C}_+ & i \mathcal{C}_+ \sin \alpha + \mathcal{C}_- \cos \alpha \\
										- i \mathcal{C}_+ \sin \alpha + \mathcal{C}_- \cos \alpha & \mathcal{C}_+
									\end{pmatrix}^{-1} ,
								\end{align}
								with ${\mathcal{C}_\pm = |a|^2 \pm |b|^2}$. The metric can be read out from the above equation:
								\begin{align}
									G(t) = \begin{pmatrix}
										\widetilde{A}_+ & - i \widetilde{A}_+ \sin \alpha + \widetilde{A}_-\\
										i \widetilde{A}_+ \sin \alpha + \widetilde{A}_- & \widetilde{A}_+
									\end{pmatrix}, \label{ConstantPTMetric}
								\end{align}
								where
								\begin{align}
									\widetilde{A}_\pm = \dfrac{|b|^2 \pm |a|^2}{2(|a|^4 + |b|^4) \cos^2 \alpha}.
								\end{align}
								Not only does this method recover the time-independent $G(t)$ that was found using the brute-force method, but it also satisfies the constraints on these coefficients.
								
								The proportional constants of the eigenvectors are chosen to be ${a = 1/\sqrt{2 \cos \alpha}}$ and ${b = i a}$ in~\cite{BenderMakingSense}. This indeed leads to 
								\begin{align}
									G(t) = \frac{1}{\cos \alpha} \begin{pmatrix}
										1 & - i \sin \alpha\\
										i \sin \alpha & 1
									\end{pmatrix}
								\end{align} provided in \cite{BenderMakingSense}.
				
							\paragraph{(2) Instantaneously Diagonal Choice}~\\
							
								In the standard choice, it is obvious that the coefficients in the constant metric, given in Eq.~\eqref{ConstantPTMetric}, could be altered by rescaling the bases. However, the metric found by the brute-force method can have time dependence, which cannot be reconstructed using the eigenvectors of the Hamiltonian as bases. It is mentioned in Sec.~\ref{InnerProductScetion} that different metrics $G(t)$ correspond to different choices of bases. Therefore, another set of bases can be introduced such that the metric has no off-diagonal parts at $t=0$, and the bases are chosen to be
								\begin{align}
									\ket{\mathbf{1}(0)} = \begin{pmatrix}
										c\\
										0
									\end{pmatrix} ~ \text{and} ~ \ket{\mathbf{2}(0)} = \begin{pmatrix}
										0\\
										d
									\end{pmatrix}.
								\end{align}
								The time evolution of the bases are
								\begin{align}
									\begin{split}
										\displaystyle \ket{\mathbf{1}(t)} = \frac{c e^{- i \gamma t}}{\cos \alpha} \begin{pmatrix}
											\cos \left( t s \cos \alpha - \alpha \right)\\
											- i \sin \left( t s \cos \alpha \right)
										\end{pmatrix},\\
										\displaystyle \ket{\mathbf{2}(t)} = \frac{d e^{- i \gamma t}}{\cos \alpha}\begin{pmatrix}
											- i \sin \left( t s \cos \alpha \right)\\
											\cos \left(t s \cos \alpha + \alpha\right)
										\end{pmatrix},
									\end{split}
								\end{align}
								where $\cos \alpha = \sqrt{1 - \left(\frac{r}{s}\right)^2 \sin^2 \theta}$ and $\gamma = \sqrt{r^2  - s^2 \sin^2 \alpha}$. Using Eq.~\eqref{AChoice}, the components of $G(t)$ are found to be:
								\begin{align}
									\begin{split}
										& g_{11} (t) = - A'_- \cos \phi_+(t) + B' \sin \phi_+(t) + A'_+,\\
										& g_{22} (t) = A'_- \cos \phi_-(t) - B' \sin \phi_-(t) + A'_+,\\
										& g_{12} (t) = - i \left( A'_- \sin \phi_0(t) + B' \cos \phi_0(t) + A'_+ \sin \alpha \right),\\
										& g_{21} (t) = g_{12}^* (t),
									\end{split}
								\end{align}
								where 
								\begin{align}
									& \phi_\pm(t) \equiv \phi_0 (t) \pm \alpha,\\
									& \phi_0(t) \equiv 2 t s \cos \alpha,\\
									& A'_\pm = \dfrac{|a|^2 \pm |b|^2}{2 |a b|^2 \cos \alpha},
								\end{align}
								and
								\begin{align}
									B' = - A_+ \tan \alpha.
								\end{align}
								The relations between $A'_\pm$ and $B'$ indeed satisfy the constraints found by the brute-force method. Since in this case $B'$ never vanishes for $\alpha \neq 0$ (the non-Hermitian case), the metric has to be time dependent. Setting $a=1=b$, the metric is $\mathbbm{1}$ at $t=0$; however, the off-diagonal parts appear as time evolves.
				
			\subsubsection{\PT-Broken Region}
			
				We now proceed to the case for $s^2 < r^2 \sin \theta$. Using the brute-force method, the components of $G(t)$ are found to be
				\begin{align}
					\begin{split}
						& g_{11}(t) = -\widetilde{\mathcal{A}}_+ \Lambda_- e^{2 \lambda t} + \widetilde{\mathcal{A}}_- \Lambda_+ e^{-2 \lambda t} + \widetilde{\mathcal{B}},\\
						& g_{22}(t) = \widetilde{\mathcal{A}}_+ \Lambda_+ e^{2 \lambda t} - \widetilde{\mathcal{A}}_- \Lambda_- e^{-2 \lambda t} + \widetilde{\mathcal{B}},\\
						& g_{12}(t) = - i \left(\widetilde{\mathcal{A}}_+ e^{2 \lambda t} + \widetilde{\mathcal{A}}_- e^{-2 \lambda t} + \widetilde{\mathcal{B}} \frac{r}{s} \sin \theta \right) + \widetilde{\mathcal{C}},\\
						& g_{21}(t) = g_{12}^* (t),
					\end{split}
				\end{align}
				where ${\lambda = \sqrt{r^2 \sin^2 \theta - s^2}}$ and ${\Lambda_{\pm} =\frac{\lambda}{s} \pm \frac{r}{s} \sin \theta  }$. The Hermiticity and positive-definiteness of $G(t)$ restrict the coefficients to obey ${\widetilde{\mathcal{A}}_\pm\frac{r}{s} \sin \theta > 0}$, ${\widetilde{\mathcal{B}} > - (\widetilde{\mathcal{A}}_+ + \widetilde{\mathcal{A}}_-) \text{sgn}\left(\frac{r }{s} \sin \theta\right)}$, ${\widetilde{\mathcal{C}} \in \mathbb{R}}$, and ${\left(4\widetilde{\mathcal{A}}_+ \widetilde{\mathcal{A}}_- - \widetilde{\mathcal{B}}^2\right)\lambda^2 > \widetilde{\mathcal{C}}^2}s^2$. In other words, the metric has to be time dependent.
				
				Next, the generalized-completeness-relation method is used to find the corresponding metric. The bases are chosen to be the eigenvectors of the Hamiltonian:
				\begin{align}
					\begin{split}
						& \ket{\mathbf{1}(t)} = a \exp(-i t r \cos \theta + t \lambda) \begin{pmatrix}
							s\\
							-i \left( r \sin \theta - \lambda \right)
						\end{pmatrix},\\
						& \ket{\mathbf{2}(t)} = b \exp(-i t r \cos \theta - t \lambda) \begin{pmatrix}
							i \left( r \sin \theta - \lambda \right)\\
							s
						\end{pmatrix},
					\end{split}
				\end{align}
				where $\lambda$ is defined above.
				
				Solving Eq.~\eqref{AChoice}, the components of $G(t)$ become
				\begin{align}
					\begin{split}
						& g_{11}(t) = -A_b \Lambda_- e^{2 \lambda t} + A_a \Lambda_+ e^{-2 \lambda t},\\
						& g_{22}(t) = A_b \Lambda_+ e^{2 \lambda t} - A_a \Lambda_- e^{-2 \lambda t},\\
						& g_{12}(t) = - i \left(A_b e^{2 \lambda t} + A_a e^{-2 \lambda t}\right),\\
						& g_{21}(t) = g_{12}^* (t),
					\end{split}
				\end{align}
				where ${A_x = \dfrac{s}{2 |x|^2 r \sin \theta}}$ for $x=a,b$. This confirms that even if the eigenvectors of the Hamiltonian are chosen as the bases, the metric is still dynamical over time.
			
			\subsubsection{Exceptional Point}
			
				There are many interesting phenomena which occur when some eigenstates coalesce. As mentioned in the beginning of this section, the Hamiltonian becomes nondiagonalizable at $s^2 = r^2 \sin \theta$. The application of the brute-force method is almost parallel to the previous examples; however, the generalized-completeness-relation method using a standard choice of bases (the eigenstates of the Hamiltonian) needs extra care. The brute-force method gives the following result:
				\begin{align}
				\begin{split}
						& g_{11}(t) = 2 r \sin \theta (\mathcal{A}' r \sin \theta + s \mathcal{B}') t^2\\
						& \quad \quad ~ - 2 \left[ r \sin \theta (\mathcal{A}' + \mathcal{C}') + s \mathcal{B}' \right] t + ( \mathcal{A}' + \mathcal{C}' ),\\
						&g_{22}(t) = 2 r \sin \theta (\mathcal{A}' r \sin \theta + s \mathcal{B}') t^2\\
						& \quad \quad ~ + 2 \left[ r \sin \theta (\mathcal{A}' - \mathcal{C}') + s \mathcal{B}' \right] t + ( \mathcal{A}' - \mathcal{C}' ),\\
						& g_{12}(t) = \mathcal{D}' - i \left[ 2 s (\mathcal{A}' r \sin \theta + s \mathcal{B}') t^2 - 2 \mathcal{C}' s t - \mathcal{B}' \right],\\
						& g_{21}(t) = g_{12}^*(t),
					\end{split} \label{EPMetricComponent}
				\end{align}
				and the constraints are all undetermined constants being real, ${\mathcal{A}'^2 - \mathcal{B}'^2 - \mathcal{C}'^2 - \mathcal{D}'^2 > 0}$, and ${\left(\mathcal{A}'^2 - \mathcal{C}'^2 \right) r^2 \sin^2 \theta > s^2 \mathcal{B}'^2}$. Like in the \PT-broken case, the metric has to be time dependent.
				
				Working with the generalized-completeness-relation method, the eigenvectors of the Hamiltonian are again treated as the bases. However, this set consists only of one eigenvector, and obviously cannot form a complete set of bases for a two-dimensional quantum space. The completion requires the set of basis vectors to include a generalized eigenvector so that the basis vectors are
				\begin{align}
					\ket{\mathbf{1}(0)} = a \begin{pmatrix}
						i \frac{r}{s} \sin \theta\\
						1
					\end{pmatrix}, ~ \ket{\mathbf{2}(0)} = \begin{pmatrix}
						a - b\\
						i b \frac{r}{s} \sin \theta
					\end{pmatrix},
				\end{align}
				where $\ket{\mathbf{1}(0)}$ is the eigenvector and $\ket{\mathbf{2}(0)}$ is the generalized eigenvector (with a rescaling so both basis vectors have the same unit). The time evolutions read
				\begin{align}
					\ket{\mathbf{1}(t)} = a \exp(-i r t \cos \theta) \begin{pmatrix}
						i \frac{r}{s} \sin \theta\\
						1
					\end{pmatrix}
				\end{align}
				and
				\begin{align}
					\ket{\mathbf{2}(t)} = \exp(-i r t \cos \theta) \begin{pmatrix}
						a r t \sin \theta + a - b\\
						i \left( - a s t + b \frac{r}{s} \sin \theta \right)
					\end{pmatrix}.
				\end{align}
				
				The result is formally the same as the components in Eq.~\eqref{EPMetricComponent} but the coefficients are now
				\begin{align}
					\begin{split}
						& \mathcal{A}' = \dfrac{1}{2|a|^4} \left[ 3 |a|^2 + 2 |b|^2 - 2 \Re (a \overline{b})\right],\\
						& \mathcal{B}' = \dfrac{-r \sin \theta}{|a|^4 s} \left[ |a|^2 + |b|^2 - \Re (a \overline{b}) \right],\\
						& \mathcal{C}' = \dfrac{1}{2|a|^4} \left[ - |a|^2 + 2 \Re(a \overline{b}) \right],\\
						& \mathcal{D}' = \dfrac{- r \sin \theta}{|a|^4 s} \Im (a \overline{b}).
					\end{split}
				\end{align}
				It can be checked that these constants indeed satisfy all the constraints obtained from the brute-force method.
		\section{Appnedix: Generalized Density Matrices in NHQM \label{AppDensityMatrices}}
		
			This Appendix covers all the subjects about the GDM, as mentioned in Sec.~\ref{DensityMatrices}, except the no-go theorems.
			
			\subsection{Pure-State Expansion of Generalized Density Matrices}
			
				The density matrices are most useful when the relative phases between the components in a given state are not known, which corresponds to a mixed state. The expectation value of an observable $\mathcal{O}$ on a system in a mixed state can still be found as follows:
			\begin{align}
				\left< \mathcal{O} \right> = \tr \left[ \rho(t) \mathcal{O} \right], \label{ExpectationOutcome}
			\end{align}
			where $\rho(t)$ is the density matrix of a given system. The derivation is split into two parts for pure and mixed states.
			
			\subsubsubsection{Pure State Case}
				By definition, a pure state is a ray in a Hilbert space, $\Ket{\phi (t)}$. The expectation of $\mathcal{O}$ is simply
				\begin{align}
					\left< \mathcal{O} \right> = \Bra{\phi (t)} \mathcal{O} \Ket{\phi (t)} = \tr \left[ \Ket{\phi (t)} \Bra{\phi (t)} \mathcal{O} \right].
				\end{align}
				Hence it is a trivial case of Eq.~\eqref{DefDensityMatrices}, where $p_1 = 1$ and $\Ket{\psi_1(t)} = \Ket{\phi(t)}$.
			
			\subsubsubsection{Mixed State Case}
				For a quantum system that the probability of obtaining $\Ket{\psi_n(t)}$ is $p_n$ for ${n=1,2,\cdots}$. Then the state can be written as
				\begin{align}
					\Ket{\psi_{\lbrace \theta \rbrace} (t)} = \sum_i \sqrt{p_i}e^{i \theta_i} \Ket{\psi_i(t)},
				\end{align}
				where $\lbrace \theta \rbrace$ is short for $\lbrace \theta_1, \theta_2, \cdots \rbrace$. Then the expectation value of the observable $\mathcal{O}$ for the state $\Ket{\psi_{\lbrace \theta \rbrace} (t)}$ is
				\begin{align}
					\begin{split}
						& \left< \mathcal{O} \right>_{\lbrace \theta \rbrace} = \left[ \sum_j \sqrt{p_j}e^{- i \theta_j} \Bra{\psi_j(t)} \right] \mathcal{O} \left[ \sum_i \sqrt{p_i}e^{i \theta_i} \Ket{\psi_i(t)} \right]\\
						& = \tr \left\lbrace \left[ \sum_i \sqrt{p_i}e^{i \theta_i} \Ket{\psi_i(t)} \right] \left[ \sum_j \sqrt{p_j}e^{- i \theta_j} \Bra{\psi_j(t)} \right] \mathcal{O} \right\rbrace.
					\end{split}
				\end{align}
				However, without any prior knowledge of the phases, the outcome of $\mathcal{O}$ can only be averaged over all phases uniformly. If we set $A_{ij} (t) \equiv \sqrt{p_i p_j} \Ket{\psi_i(t)} \Bra{\psi_j(t)}$, the average outcome of the ensemble becomes
				\begin{align}
					\begin{split}
						& \left< \mathcal{O} \right> = \int_{0}^{2 \pi} \prod_{k} \frac{d\theta_{k}}{2 \pi} \left< \mathcal{O} \right>_{\lbrace \theta \rbrace}\\
						& \quad = \int_{0}^{2 \pi} \prod_{k} \frac{d\theta_{k}}{2 \pi} \tr \left[ \sum_{i,j} e^{i (\theta_i - \theta_j)} A_{ij} (t) \mathcal{O} \right]\\
						& \quad = \int_{0}^{2 \pi} \prod_{k} \frac{d\theta_{k}}{2 \pi} \tr \left[ \sum_{i} A_{ii} (t) \mathcal{O} \right]\\
						& \quad \quad ~ + \int_{0}^{2 \pi} \prod_{k} \frac{d\theta_{k}}{2 \pi} \tr \left[ \sum_{i \neq j} e^{i (\theta_i - \theta_j)} A_{ij} (t) \mathcal{O} \right]\\
						& \quad = \tr \left[ \sum_{i} A_{ii} (t) \mathcal{O}\right]\\
						& \quad = \tr \left[ \sum_{i} p_i \Ket{\psi_i(t)} \Bra{\psi_i(t)} \mathcal{O} \right],
					\end{split}\label{EnsembleAverage}
				\end{align}
				where the second to the last equality is a result of $\int d\theta \exp(i\theta) = 0$. Comparison of Eq.~\eqref{EnsembleAverage} and Eq.~\eqref{ExpectationOutcome} shows that
				\begin{align}
					\rho = \sum_i p_i \Ket{\psi_i(t)} \Bra{\psi_i(t)}, \label{DefDensityMatrices2}
				\end{align}
				indeed, plays the role of a density matrix. The rest of this Appendix is to show that the GDM satisfy the postulates of a true density matrix and its time evolution.
			
		\subsection{Self-Adjointness and Positive Semi-Definiteness of Generalized Density Matrices}
			
			It is well known that a density matrix has to be self-adjoint, positive semidefinite, and the trace being unity. Here we show that the density matrix, defined in Eqs.~\eqref{DefDensityMatrices} and Eq.~\eqref{DefDensityMatrices2}, is self-adjoint and positive semidefinite. To demonstrate that the density matrix is self-adjoint, it is useful to find the Hermitian conjugate of $\rho (t)$:
			\begin{align}
				\begin{split}
					& \rho^\dagger (t) = \left[ \sum_i p_i \ket{\psi_i(t)} \bra{\psi_i(t)} G(t) \right]^\dagger\\
					& \quad = G(t) \sum_i p_i \ket{\psi_i(t)} \bra{\psi_i(t)}\\
					& \quad = G(t) \sum_i p_i \ket{\psi_i(t)} \bra{\psi_i(t)} G(t) G^{-1}(t)\\
					& \quad = G(t) \rho(t) G^{-1}(t).
				\end{split}\label{DiracHermitianDensityMatrix}
			\end{align}
			Acting with $G(t)$ on both sides of Eq.~\eqref{DiracHermitianDensityMatrix} from the right gives
			\begin{align}
				\rho^\dagger (t) G(t) = G(t) \rho(t).
			\end{align}
			By the self-adjoint condition, given in Eq.~\eqref{HermitianOperators}, it is clear that the density matrix is indeed self-adjoint.
			
			Showing that the density matrix is positive semidefinite is quite straightforward:
			\begin{align}
				& \Bra{\Psi(t)} \rho(t) \Ket{\Psi(t)} = \sum_i p_i \Braket{\Psi(t)}{\psi_i(t)} \Braket{\psi_i(t)}{\Psi(t)} \nonumber \\
				& \quad = \sum_i p_i \left| \Braket{\psi_i(t)}{\Psi(t)} \right|^2 \geq 0,
			\end{align}
			where the second equality comes from Eq.~\eqref{ComplexConjugate}. To show that the trace of this density matrix is unity at every instant, it is necessary to find the dynamics of $\rho(t)$ which is discussed in the next part of this Appendix.
			
		\subsection{Time Evolution of Generalized Density Matrices}
		
			The time evolution of every states is governed by Schr\"{o}dinger's equation
			\begin{align}
				\partial_t \Ket{\psi_i (t)} = - i H(t) \Ket{\psi_i (t)}.
			\end{align}
			Since the corresponding bra is defined as $\Bra{\psi_i (t)} = \bra{\psi_i (t)} G(t)$, the time evolution can be found by using the Leibniz rule:
			\begin{align}
				\begin{split}
					& \partial_t \Bra{\psi_i (t)} = [\partial_t \bra{\psi_i (t)}] G(t) + \bra{\psi_i (t)} \partial_t G(t)\\
					& \quad = [ i \bra{\psi_i (t)} H^\dagger (t) ] G(t)\\
					& \quad \quad ~ +  \bra{\psi_i (t)} \left[ i G(t) H(t) - i H^\dagger (t) G(t) \right]\\
					& \quad = i \bra{\psi_i (t)} G(t) H(t) = i \Bra{\psi_i (t)} H(t).
				\end{split}
			\end{align}
			Inserting the above result into Eq.~\eqref{EnsembleAverage} or Eq.~\eqref{ExpectationOutcome} implies
			\begin{align}
				\begin{split}
					\partial_t \rho(t) & = \partial_t \sum_i p_i \Ket{\psi_i(t)} \Bra{\psi_i(t)}\\
					& = \sum_i p_i \left[ \partial_t \Ket{\psi_i(t)} \right] \Bra{\psi_i(t)}\\
					& \quad ~ +  \sum_i p_i \Ket{\psi_i(t)} \left[ \partial_t \Bra{\psi_i(t)}\right]\\
					& = \frac{- i}{\hbar} \sum_i p_i H \Ket{\psi_i(t)} \Bra{\psi_i(t)}\\
					& \quad ~ + \frac{i}{\hbar}  \sum_i p_i \Ket{\psi_i(t)}  \Bra{\psi_i(t)}H\\
					& = - \frac{i}{\hbar} [ H , \rho (t) ].
				\end{split} \label{DensityMatrixDynamics}
			\end{align}
			Therefore, the GDM obeys the quantum Liouville equation for any Hamiltonian.
			
		\subsection{The Trace of Generalized Density Matrices}
		
			Equation \eqref{DensityMatrixDynamics} is almost a solid evidence that the trace of a density matrix is invariant under time evolution. The only missing piece is to show that the trace commutes with the time derivative. To prove this, the trace is also modified to
			\begin{align}
				\tr (A(t)) = \sum_n \Bra{n(t)} A(t) \Ket{n(t)}.
			\end{align}
			The reason for writing explicitly the time dependence in this equation is to make sure that the complete set of states also evolves with the operator. Before showing that the time derivative commutes with the trace operation, it is useful to show the cyclic property:
			\begin{align}
				\begin{split}
					& \tr [ A(t) B(t) C(t) ] = \sum_n \Bra{n(t)} A(t) B(t) C(t)\Ket{n(t)}\\
					& \quad = \sum_{m,n} \Bra{n(t)} A(t) \Ket{m(t)} \Bra{m(t)} B(t) C(t) \Ket{n(t)}\\
					& \quad = \sum_{m,n} \Bra{m(t)} B(t) C(t) \Ket{n(t)} \Bra{n(t)} A(t) \Ket{m(t)}\\
					& \quad = \sum_m \Bra{m(t)} C(t) B(t) A(t) \Ket{m(t)}\\
					& \quad = \tr [ B(t) C(t) A(t) ],
				\end{split}\label{CyclicTrace}
			\end{align}
			where the generalized completeness relation, given in Eq.~\eqref{CompletenessRelation}, is used repeatedly.
			
			With the cyclic property of the trace, it is easy to prove that the trace and time derivative operations commute:
			\begin{align}
				\begin{split}
					& \tr \left[ \partial_t A(t) \right] - \partial_t \tr \left[ A(t) \right] \\
					& = \sum_m \Bra{m(t)} \left[ \partial_t A(t) \right] \Ket{m(t)} - \partial_t \sum_n \Bra{n(t)} A(t) \Ket{n(t)}\\
					& = -\sum_n \big\lbrace \left[ \partial_t \Bra{n(t)} \right] A(t) \Ket{n(t)} + \Bra{n(t)} A(t) \left[ \partial_t \Ket{n(t)} \right] \big\rbrace\\
					& = \frac{i}{\hbar} \sum_n \big\lbrace \Bra{n(t)} [ H(t) A(t) - A(t) H(t)] \Ket{n(t)} \big\rbrace = 0,
				\end{split}
			\end{align}
			where the last equality uses the cyclic property of the trace operation.
			
			With these useful properties of the trace operation, Eq.~\eqref{DensityMatrixDynamics} shows that the trace of a density matrix is constant in time:
			\begin{align}
				\tr \left[ \partial_t \rho(t) \right] = - \frac{i}{\hbar} \tr [ H , \rho (t) ] \Rightarrow \partial_t \tr \left[ \rho(t) \right] = 0,
			\end{align}
			where the cyclic property of the trace operation is used. When $\Braket{\psi_i (t)}{\psi_i (t)} = 1$ for all $i$ and $\displaystyle \sum_i p_i = 1$, one observes that
			\begin{align}
				& \tr \rho(t) = \sum_i p_i \Braket{\psi_i (t)}{\psi_i (t)} = \sum_i p_i = 1.
			\end{align}
			This result finally shows that the GDM qualifies to be a density matrix in the Hilbert space.

	\bibliography{references}

\end{document}